\documentclass[10pt,journal,compsoc]{IEEEtran}

\ifCLASSOPTIONcompsoc
  \usepackage[nocompress]{cite}
\else
  \usepackage{cite}
\fi

\ifCLASSOPTIONcompsoc
  \usepackage[caption=false,font=footnotesize,labelfont=sf,textfont=sf]{subfig}
\else
  \usepackage[caption=false,font=footnotesize]{subfig}
\fi
\ifCLASSOPTIONcompsoc
  \usepackage[nocompress]{cite}
\else
  \usepackage{cite}
\fi
\usepackage{url}
\usepackage[pdftex]{graphicx}
\usepackage{algorithmic}
\usepackage{colortbl}
\usepackage{etoolbox}
\usepackage{tikz}
\usepackage{amsmath}

\graphicspath{{figures/} {figures_tex/} {figures_final/}}
\definecolor{lightgray}{gray}{0.9}
\definecolor{white}{gray}{1}

\begin{document}

\author{Dennis~Pinto,
        Jose-María~Arnau,
        and~Antonio~González,~\IEEEmembership{Fellow,~IEEE}
\IEEEcompsocitemizethanks{\IEEEcompsocthanksitem D. Pinto, JM. Arnau and A. González are with the Department of Computer Architecture, Universitat Politècnica de Catalunya, Barcelona, Spain.\protect\\
E-mail: \{dpinto, jarnau, antonio\}@ac.upc.edu}
}

\title{Mixture-of-Rookies: Saving DNN Computations by Predicting ReLU Outputs}

\IEEEtitleabstractindextext{%

\begin{abstract} 
Deep Neural Networks (DNNs) are widely used in many applications domains. However, they require a vast amount of computations and memory accesses to deliver outstanding accuracy. In this paper, we propose a scheme to predict whether the output of each ReLu activated neuron will be a zero or a positive number in order to skip the computation of those neurons that will likely output a zero. Our predictor, named \textit{Mixture-of-Rookies}, combines two inexpensive components. The first one exploits the high linear correlation between binarized (1-bit) and full-precision (8-bit) dot products, whereas the second component clusters together neurons that tend to output zero at the same time. We propose a novel clustering scheme based on analysis of angles, as the sign of the dot product of two vectors depends on the cosine of the angle between them. We implement our hybrid zero output predictor on top of a state-of-the-art DNN accelerator. Experimental results show that our scheme introduces a small area overhead of 5.3\% while achieving a speedup of 1.2x and reducing energy consumption by 16.5\% on average for a set of diverse DNNs.
\end{abstract}
\begin{IEEEkeywords}
neural networks, energy efficiency, automatic speech recognition, hardware acceleration
\end{IEEEkeywords}}

\maketitle
\IEEEdisplaynontitleabstractindextext
\IEEEpeerreviewmaketitle

\section{Introduction}
\textit{Deep Neural Networks (DNNs)} are the most successful and widely used family of machine learning methods. DNNs deliver state-of-the-art performance for tasks such as speech recognition~\cite{pratap2020scaling} or image classification~\cite{he2016deep}. However, high recognition accuracy comes at the cost of using large DNNs that require a vast amount of computations and memory accesses~\cite{akhlaghi2018snapea}. For example, classifying a single image from \textit{ImageNet}~\cite{ImageNetDeng} requires billions of multiply-and-accumulate operations and memory accesses~\cite{he2016deep}.

\begin{figure}[t]
\centering
\includegraphics[width=0.8\linewidth]{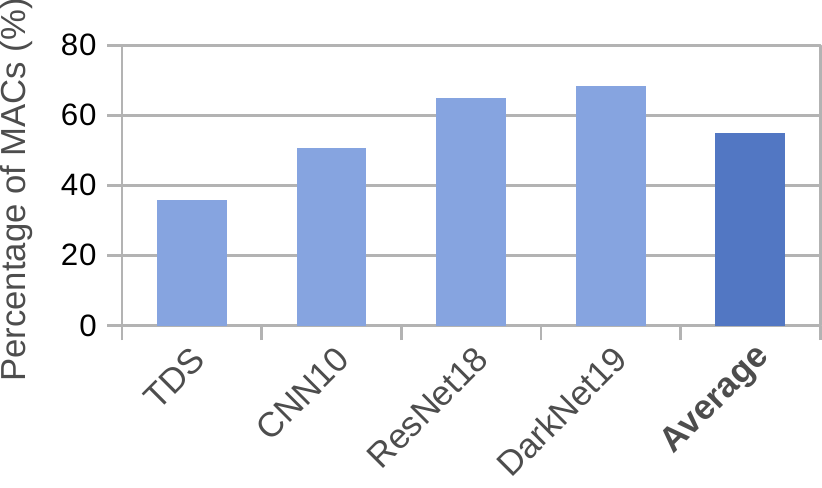}
\caption{Percentage of computations performed to produce negative ReLU inputs. On average, 55\% of the computations
produce negative inputs that are turned into zeros by ReLU activation functions.}
\label{f:motivation}
\end{figure}

The vast majority of computations occur in \textit{fully-connected (FC)} and \textit{convolutional (CONV)} layers~\cite{MarcComputationReuse}. FC and CONV layers are commonly followed by an activation function called \textit{Rectifying Linear Unit (ReLU)}~\cite{AlexNet} that returns the unmodified input value for positive inputs and zero for negative inputs. Previous work observed that a large fraction of ReLU inputs are negative values~\cite{akhlaghi2018snapea}. Our own results collected on more recent DNNs ratify this claim. Figure~\ref{f:motivation} shows that between 35\% and 69\% of the computations generate negative inputs that are converted into zeros by the ReLU activation function. Therefore, if we could predict in advance which neurons will have negative ReLU inputs in FC/CONV layers, we could set their outputs to zero without any computation and save 55\% of the operations on average.

In this paper, we propose a zero output predictor for FC and CONV layers with ReLU activation functions. Computing each output in an FC or CONV layer requires a dot product, usually performed as a sequence of \textit{Multiply and Accumulate (MAC)} operations, between a vector of inputs and a vector of weights. The result of the dot product is the input to the ReLU activation function. The objective of the predictor is to determine the sign of this dot product. If the predictor indicates that the result will be positive, then the corresponding neuron is evaluated as normal. Otherwise, the computation of the dot product is skipped, avoiding all the associated computations and memory fetches, and the output of the neuron is simply set to zero.

The predictor must be accurate and low-cost. Incorrect predictions may introduce errors in the DNN, potentially decreasing its accuracy. More specifically, incorrectly predicting a dot product outcome as negative introduces an error, whereas incorrectly predicting a dot product as positive has no penalty in accuracy but represents a missed opportunity to save computations. On the other hand, this scheme is only beneficial if the overhead of the predictor is lower than the cost of computing the dot product. Prior negative input value predictors~\cite{lin2017predictivenet,cao2019seernet,akhlaghi2018snapea} introduce a significant overhead that severely impact the speedups and energy savings. Hence, the key challenge is to design a predictor that is low-cost and highly accurate.

Our solution consists of a hybrid predictor, \textit{Mixture-of-Rookies}, composed of two low-cost components to accurately identify negative ReLU inputs with inexpensive hardware. 

The first component is based on the observation that the output of the \textit{binarized} dot product, i.e. after converting inputs and weights to 1-bit, shows high linear correlation with the original 8-bit dot product~\cite{silfa2019neuron}. For this purpose, we use a subset of the training set to perform a linear regression between binarized and 8-bit outputs, obtaining a fitted line for each neuron/filter. During inference, we compute first the binarized version of the dot product, and use the fitted line to estimate the output value of the full-precision dot-product. If the estimated value is negative, we predict that the result of the dot-product in base precision will also be negative so the outcome of the neuron (after ReLU) will be zero, and the corresponding computations and memory accesses can be avoided. We show that the inexpensive hardware required to support binary dot-products results in very low overhead over a state-of-the-art DNN accelerator.

The second component takes a different approach as it exploits correlation among neurons in the same FC/CONV layer. We observe that there are groups of neurons whose dot-product have the same sign for the majority of the executions. In other words, if one element of the group produces a negative dot-product result, the rest of the neurons tend to produce a negative result, too. For this predictor, we first find these groups of highly correlated neurons, and then evaluate one representative element from each group; if the representative neuron has a negative ReLU input, it predicts that the output for all the neurons of the group will be zero.

Finding a minimum set of representative elements is the main challenge for the latter predictor. To this end, we leverage information about the angles between weight vectors. We observe that the sign of the dot product depends exclusively on the cosine of the angle between input and weight vectors.

First, we only group neurons that share the same input vector. Furthermore, if two neurons show a small angle between their weight vectors, it is very likely that the dot product with the same input vector will exhibit the same sign for both neurons. Therefore, we propose a novel mechanism that clusters neurons together by analyzing the angle among their weight vectors. We show that this predictor has negligible cost at run-time, since the clustering is performed only once offline, as the weights are fixed for a given model.

Combining the above two predictors, the proposed hybrid scheme predicts that the output of a neuron will be zero if and only if both components
indicate so. More specifically, first one representative element from each group is fully computed. For those that yield a 
negative ReLU input, we evaluate the binarized dot product for the remaining elements in
the group, and use the fitted line to estimate full-precision ReLU input. If the estimated value is 
also negative, we skip the computation of the dot product for that neuron and predict the output as zero. The hybrid
predictor avoids 18\% of computations and on-chip memory accesses and saves 17\% of main
memory traffic on average for a diverse set of DNNs. This results in 1.2x speedup and
16.5\% energy savings with a small area overhead of 5.3\%. We also show that the hybrid predictor yields much better results than any of 
its two components in isolation.

Note that many DNNs include a batch normalization~\cite{pmlr-v37-ioffe15} between
the dot product and the ReLU activation function. We also describe in this paper how our hybrid predictor
can be applied in the presence of batch normalization.

To sum up, the main contributions of this work are the following:
\begin{itemize}
    \item We propose a neuron output predictor for any neuron that uses a ReLU activation function, which is based on exploiting two main properties: a) the high correlation among some neurons within the same layer, and b) the high correlation between the output of a neuron and the output of the same neuron when inputs and weights are binarized.
    \item We show that the correlation between two neurons that have the same input vector can be accurately and efficiently estimated from the angle between their respective weight vectors.
    \item We propose the design of a novel hybrid predictor, which we call Mixture-of-Rookies, that combines the above two strategies in a synergistic manner.
    \item We evaluate Mixture-of-Rookies on top of a modern DNN accelerator. Our results show that it provides 1.2x speedup and 16.5\% energy savings.
\end{itemize}


\section{Background on ReLU Output Prediction}\label{sec:background}

Some previous works explored the use of ReLU output predictors to save computations
in DNNs. These proposals can be classified in three categories. The first type of
works exploit self-correlation between full-precision neurons and aggressively
quantized versions. The second type exploits spatial correlation among neurons in the
same layer. The last class evaluates a subset of connections to estimate the
outcome of the entire neuron. In this section, we describe the most relevant proposals in this area and highlight their main weaknesses/areas for improvement.

\subsection{ReLU Output Prediction Based on Self-correlation}\label{sec:ofmap_self}
For the rest of this work, \textit{neuron self-correlation} is defined as the degree of correlation between the output of a neuron computed in base precision (typically 8 or 16 bits) and an alternative output for the same neuron computed in lower precision, i.e. using fewer bits. Since self-correlation is generally expected to be high, it can be leveraged to predict negative outputs by computing first the dot product of the input and weight vectors in low-precision. If it is negative, the output of the neuron is predicted to be zero, and thus, all computations related to that output can be skipped. Otherwise, the output is computed in base precision. The challenge in this approach is to chose a low-precision representation that maximizes self-correlation with minimum overhead.

The literature contains several proposals that exploit self-correlation to predict negative values in ReLU layers. The authors of \textit{PredictiveNet}~\cite{lin2017predictivenet} propose to break down the inputs and weights in two halves: one containing the most significant bits and the other containing the less significant bits. The dot-products are then performed in two steps. First, a dot-product is performed with the most significant bits of each input and weight. If the result is negative, the output of the neuron is predicted to be zero and skipped. Otherwise, another dot-product is performed with the other half of each input and weight and the results are merged to obtain the final output in base precision.

A similar idea is explored in~\cite{song2018prediction}, but with some refinements. In that work, a number of most significant bits is also used to predict the outcome of ReLU networks, but following an approach similar to \textit{stripes}~\cite{judd2016stripes}. DNNs are profiled in order to obtain the minimum number of high-order bits required by each layer to predict ReLU outputs without accuracy loss. Finally, they propose an architecture based on independent dot-product units with an array of bit-serial multipliers to deal with variable precision.

Another approach is presented in \textit{SeerNet}~\cite{cao2019seernet}. This technique uses a highly quantized version of the original DNN, without retraining, to predict whether the ReLU inputs will be positive or negative. In this proposal, the dot-products are first performed with the quantized neurons to obtain an approximated ReLU input. If it is positive, the base-precision dot-product is computed to obtain the final output. Even though they consider any number of bits for the predictor, their analysis shows that predictors with less than 4-bits incur in many prediction errors, and hence, they use a 4-bit predictor obtained by linear quantization of the original DNN.

\subsection{ReLU Output Prediction Based on Spatial Correlation} \label{sec:ofmap_spatial}
Neurons in a convolutional layer have been observed to exhibit high \textit{spatial correlation}, meaning that close-by neurons tend to have similar outputs. An example of such approach includes the work performed by Shomron et al.~\cite{shomron2018spatial}. They observe that for a particular CNN application, $66\%$ of the zero-valued outputs are contained within 2x2 all-zero non-overlapping windows. The paper proposes to divide the output feature maps in non-overlapping square windows and compute first the values in the diagonals. If they are all zero, the remaining values within the window are predicted to be zero, as well. Otherwise, they are computed.

Another representative work by the same authors~\cite{shomron2020thanks} follows a different approach to exploit spatial correlation. They design a small CNN, called \textit{ZAP}, to predict whether individual ReLU inputs in convolutional layers will be positive or negative. In their proposal, a subset of the output is computed following a fixed pattern, e.g. a chessboard pattern. Then, this partial output is evaluated with ZAP to predict whether each of the non-computed outputs values should be computed or can be kept as zero.

\subsection{ReLU Output Prediction Based on Sub-sampling}

Some networks exhibit particular properties that can be exploited for early detection of negative values in convolutions, based on evaluating just a subset of the connections. A recent example is \textit{SnaPEA}~\cite{akhlaghi2018snapea}. This technique leverages the observation that some convolutional layers only have positive inputs. Their proposal consists of sorting the weights of convolutional kernels in descending order and compute the MACs for each convolution sequentially, and stop the computation for convolutions when the accumulated value becomes negative. Since all the inputs are positive, and the weights are sorted in descending order, when a ReLU input becomes negative, it will remain negative. At that point, the final input can be safely assumed to be negative, and the remaining computations can be skipped without degrading the accuracy of the DNN. To further increase the benefits, the authors propose a predictive mode in which each filter from the neural network is statically profiled to obtain a threshold, which is used to speculatively stop the computation of the convolution, instead of waiting for the accumulation to become negative.

\subsection{Discussion of Weaknesses}

Accurate predictions are key to prevent DNN accuracy loss. Achieving high precision with low implementation cost is extremely challenging. Previous solutions introduce a significant overhead to accurately identify zero ReLU outputs. For example, \textit{PredictiveNet}~\cite{lin2017predictivenet} and \textit{SeerNet}~\cite{cao2019seernet} require complete evaluation of a neuron at low precision, typically 4 bits, which represents a significant overhead given the 8-bit baseline. On the other hand, work in~\cite{shomron2020thanks} requires evaluating an entire CNN to predict a ReLU output, whereas \textit{SnaPEA}~\cite{akhlaghi2018snapea} may need to evaluate a large number of connections before identifying a negative result, while requiring complex hardware to recover weight ordering.

Recognizing the complexity of designing a low-cost and accurate predictor, in this work we take a different approach. We propose \textit{Mixture-of-Rookies}, a combination of two novel and simple ReLU output predictors, and show that it results in an accurate and effective solution with negligible overhead. Our results show that the two schemes complement each other, as one predictor is able to catch mistakes made by the other and vice versa.
\section{Low-Overhead ReLU Output Predictor} \label{sec:technique}

In this section, we first analyze how ReLU activation functions are used in modern DNNs and characterize the potential benefits of predicting ReLU outputs. We then present our ReLU output predictor that exploits two properties: (1) correlation between binarized, i.e. 1-bit inputs and weights, and base precision neurons, and (2) correlation among neurons with the same input vector.

\subsection{ReLU Activations in DNNs}

\begin{figure*}[t]
     \centering
     \subfloat[Building block for TDS neural network.]{
        \label{f:block_diagram_tds}
        \includegraphics[width=0.8\textwidth]{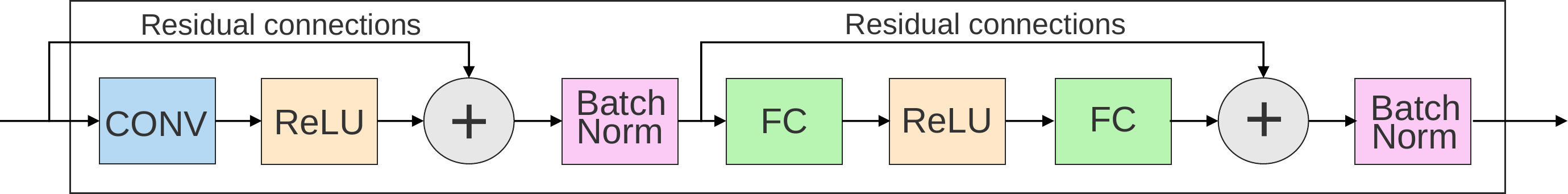}
    }
    
    \subfloat[Building block for convolutional layer.]{
        \label{f:block_diagram_conv}
        \includegraphics[width=0.2\textwidth]{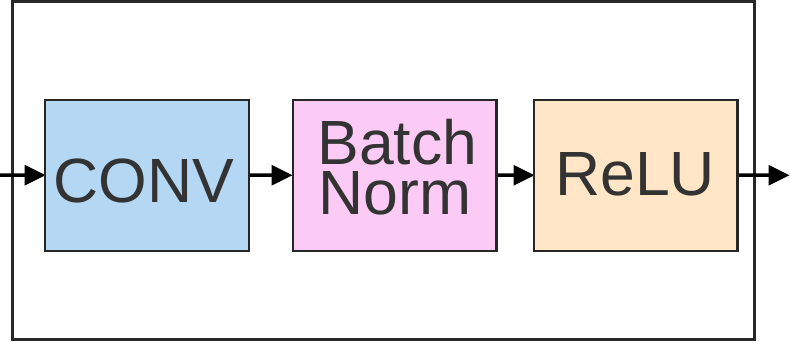}
    }
    \subfloat[Building block for Resnet.]{
        \label{f:block_diagram_resnet}
        \includegraphics[width=0.5\textwidth]{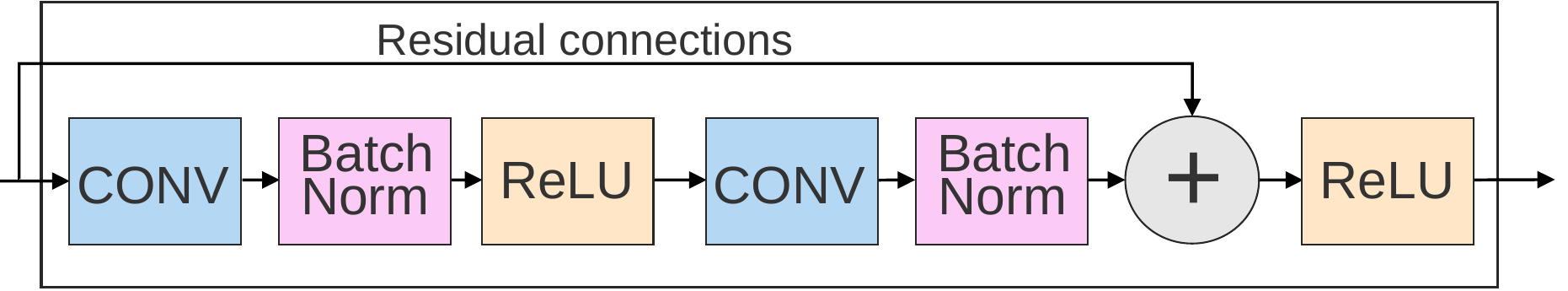}
    }    
    \caption{Building blocks for different DNNs.}
    \label{f:block_diagrams_dnns}
\end{figure*}

ReLU is probably the most popular activation function. However, DNNs differ in the way they compute ReLU inputs as shown in Figure~\ref{f:block_diagrams_dnns}. The building block of a \textit{Time-Depth Separable (TDS)}~\cite{hannun2019sequence} convolution is depicted in Figure~\ref{f:block_diagram_tds}. TDS delivers state-of-the-art accuracy for speech recognition~\cite{synnaeve2019end}. It consists of one CONV and one FC layer, both with ReLU activations, followed by another FC layer without ReLU. In a TDS block, each ReLU input is the result of a dot product between a vector of weights and a vector of inputs.

On the other hand, Figure~\ref{f:block_diagram_conv} shows the building block used by many \textit{Convolutional Neural Networks (CNNs)}. As it can be seen, before applying the activation function, ReLU inputs are \textit{batch normalized}~\cite{pmlr-v37-ioffe15} as follows:

$$
ReLU input = \frac{dotprod(\overrightarrow{weights}, \overrightarrow{inputs}) - \mu}{\sigma} * \gamma + \beta
$$

where $\mu$ and $\sigma$ are the mean and standard deviation of each dot product in the training dataset, whereas $\gamma$ and $\beta$ are learnable parameters. Another popular CNN architecture is \textit{ResNet}, whose building block is illustrated in Figure~\ref{f:block_diagram_resnet}. In addition to batch normalization, the ResNet building block includes a \textit{residual connection} before ReLU. Both batch normalization and the residual connection may change the sign of the ReLU input and, hence, they must be considered to determine whether the ReLU activation will output a zero or not.

\begin{figure}[t]
\centering
\includegraphics[width=0.8\columnwidth]{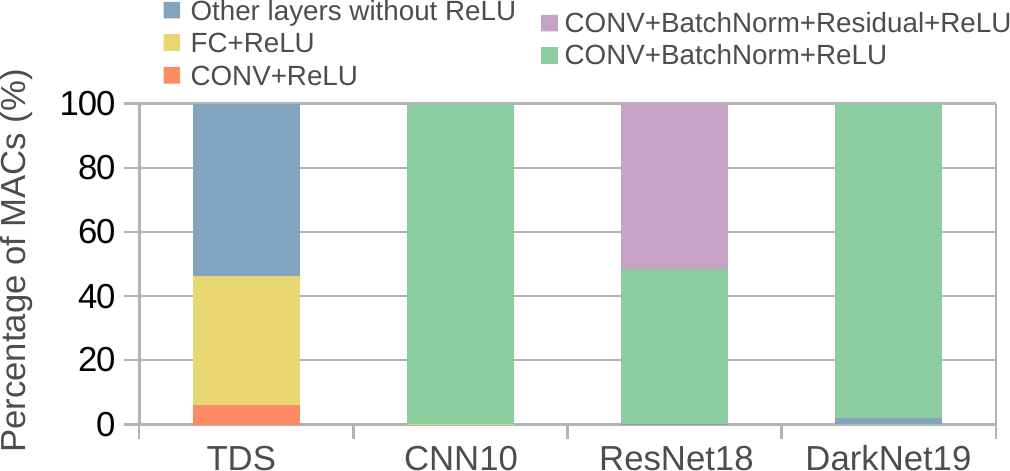}
\caption{Percentage of MACs in each type of layer for a set of DNN aplications.}
\label{fig:tds_op_breakdown}
\end{figure}

Figure~\ref{fig:tds_op_breakdown} shows the percentage of Multiply and Accumulate (MAC) operations in each type of layer for different DNNs. In the TDS model for speech recognition, CONV and FC layers with ReLU represent 6\% and 40\% of the operations respectively. Hence, a ReLU output predictor can save up to 46\% of the computations. On the other hand, more than 98\% of the computations are in CONV layers with batch normalization and ReLU for \textit{Darknet19} and \textit{CNN10} (A simple CNN composed of 10 layers as described in figure \ref{f:block_diagram_conv}). Finally, in \textit{Resnet18}, CONV layers with batch normalization and ReLU represent 48\% of the computations, whereas 52\% of MACs are performed in CONV layers that also include residual connections. Therefore, to be widely applicable, a ReLU output predictor must support CONV and FC layers and provide accurate predictions in the presence of batch normalization and residual connections.

\subsection{ReLU Output Predictor} \label{sec:binary_predictor}

In this work, we propose \textit{Mixture-of-Rookies}, a neuron output prediction scheme useful for neurons with a ReLU activation function, that is based on a combination of predictors with negligible overhead. This scheme exploits the synergies between different sources of information, improving prediction accuracy. 

More specifically, we first leverage self-correlation by binarizing the DNN and using the binarized network to predict the outcome of the neurons, and then we exploit spatial correlation with a novel approach that generates clusters of neurons according to the angle between their weight vectors. An advantage of this approach is that the predictor is not based on any property specific to a class of DNNs, instead, it is a general technique applicable to a wide range of them.

\textit{Mixture-of-Rookies} consists of 2 stages. First an offline stage performs two tasks: a) profiles the self-correlation of neurons, and b) generates clusters by grouping together neurons that share the same inputs and have the property that for any given input vector, either all the neurons in a cluster will produce a zero output or all of them will produce a non-zero output. Second, an online stage performs value prediction for the neurons during inference to avoid computing neurons whose ReLU activation function is predicted to produce a zero value.

Regarding the offline tasks, our technique employs a subset of training samples to perform a linear regression between binarized and base precision dot products in CONV and FC layers, obtaining a fitted line for each neuron. Besides, it groups the neurons of the same layer based on the similarity property described above and selects one neuron from each cluster to represent the whole group.

During DNN inference, \textit{Mixture-of-Rookies} evaluates first the representative neuron for each group at base precision. If it generates a zero ReLU output, all the other neurons in the group are evaluated using 1-bit inputs and weights and the fitted line for each neuron is used to estimate the base precision ReLU output. If the estimated output of a neuron using this approach is also zero, then all the computations and memory accesses for this neuron are skipped and its output is set to zero. Otherwise, the neuron is computed using the base precision. In other words, a neuron ReLU output will be predicted to be zero if and only if both prediction schemes agree on that. The next subsections provide further details on our predictor.

\subsubsection{Exploiting Self-correlation} \label{sec:self_correlation_pred}

Our \textit{Mixture-of-Rookies} predictor exploits linear correlation between the ReLU input of a neuron computed in full precision and the ReLU input of a binarized version of the same neuron. A neuron can be binarized by using the sign bit of its weights and inputs as described~in\cite{BinaryNet}. Exploiting this correlation, we build a predictor of each neuron's output by computing the dot-product of its binary weight and input vectors, and predicting the dot-product of these vectors in full precision (i.e., the ReLU input) using this correlation. This approach has several advantages. First, the dot-product between 1-bit valued vectors does not require multipliers, simplifying the hardware by a large extent. Second, since the 1-bit weights are obtained from the sign bits of the full precision weights~\cite{silfa2019neuron}, they do not incur in any memory footprint overhead since they do not have to be stored separately, but can be obtained directly from the full precision weights.

\begin{figure}[t]
\centering
\includegraphics[width=1\columnwidth]{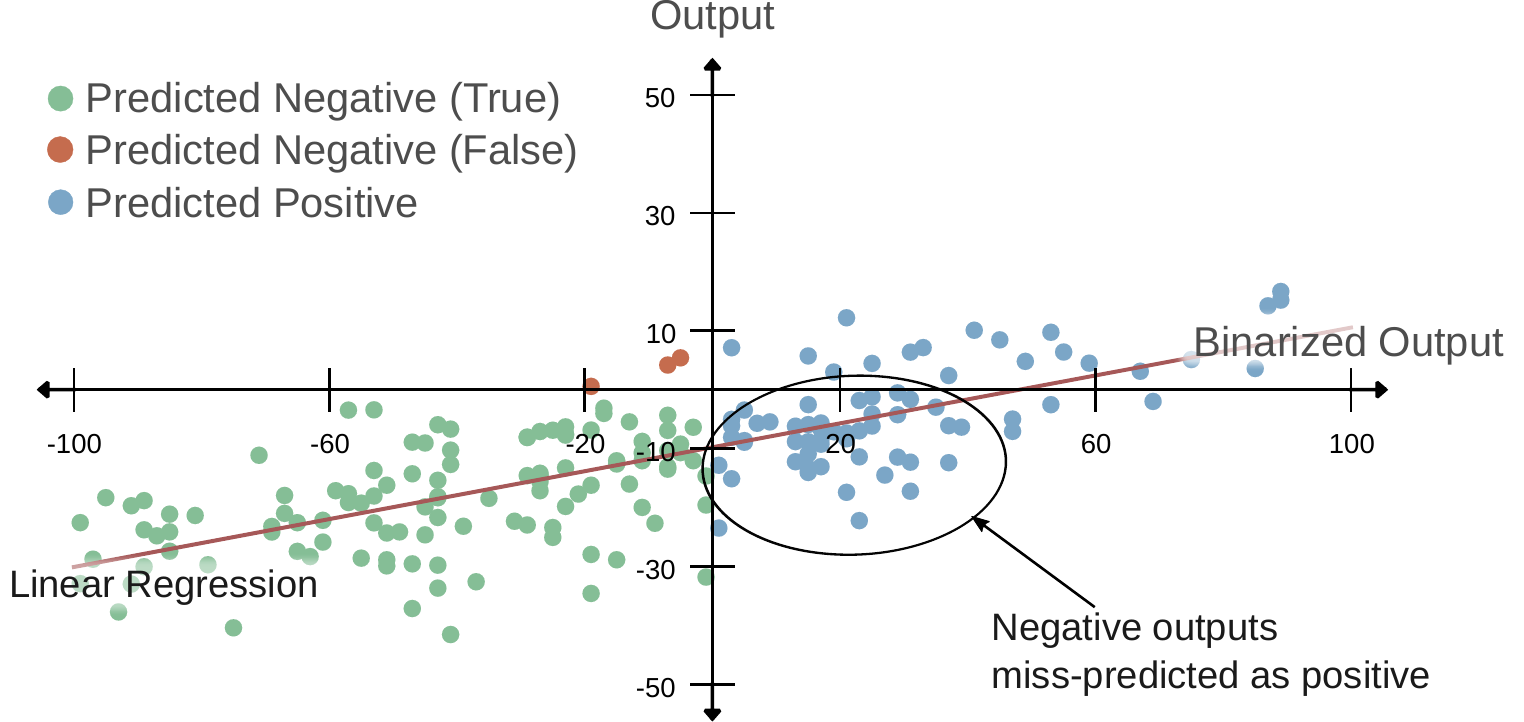}
\caption{ReLU inputs for binarized neuron (x-axis) versus ReLU inputs for base precision neuron (y-axis).}
\label{fig:correlation_nofix}
\end{figure}

Figure~\ref{fig:correlation_nofix} shows the ReLU inputs for a sample neuron from the TDS DNN in base (8-bit) precision (y-axis), versus the ReLU inputs for the binarized version of the neuron (x-axis). As it can be seen, there is a high linear correlation (correlation factor of $0.78$). However, the sign of the ReLU input for 1-bit cannot be used as an estimation of the sign for the ReLU input in base precision, as high linear correlation does not imply that the signs match. For example, points in the bottom-right quarter of Figure~\ref{fig:correlation_nofix} are positive in binarized version but negative in base precision. To mitigate this problem, we perform a linear regression and use the fitted line to obtain an estimated ReLU input from the binarized value. That is, we compute the coefficients of this fitted line and use it to transform the output of the binarized dot-product into the expected output of the base-precision dot-product.

\begin{figure}[t]
\centering
\includegraphics[width=0.8\columnwidth]{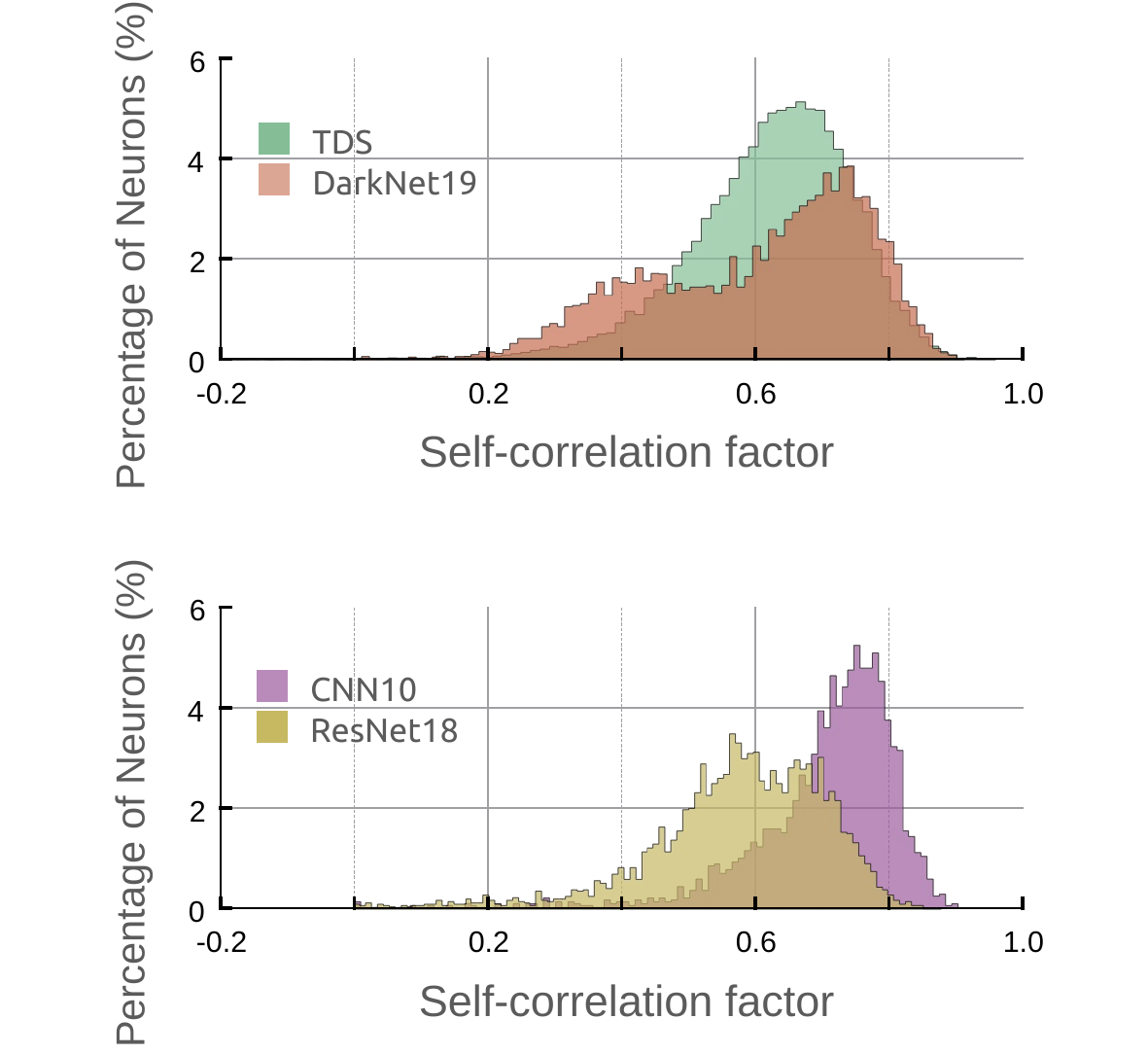}
\caption{Distribution of neurons according to the Pearson correlation coefficient of the binary and base-precision ReLU inputs.}
\label{fig:fp_bin_correlation}
\end{figure}

Figure~\ref{fig:fp_bin_correlation} shows the distribution of different levels of correlation among neurons for our bechmarks. Even though most neurons exhibit a high correlation, a significant number of neurons have moderate or even low correlations. This observation is consistent with the observations made by Anderson et al.~\cite{anderson2017high} and more recently, Silfa et al.~\cite{silfa2019neuron}. A predictor based on 1-bit weights for a neuron with a low self-correlation coefficient is expected to make frequent mistakes, and consequently, reduce the overall accuracy of the DNN. Therefore, our predictor scheme is only enabled for neurons that show high linear correlation with their binarized versions.

\textit{Mixture-of-Rookies} performs a pre-processing stage on the trained model to extend each neuron's parameters with three additional ones: correlation coefficient ($c$), slope ($m$) and y-intercept ($b$) of the fitted line. These parameters are computed by using a randomly selected subset of the training dataset. Using this training subset, for each neuron we obtain two series of data: ReLU inputs at 8-bit and 1-bit precision. We then compute the \textit{Pearson correlation factor} ($c$) between the two series and perform a linear regression to obtain a fitted line $y = mx + b$. Parameters $c$, $m$ and $b$ are saved in the DNN together with the weights.

During DNN inference, each neuron is processed as follows. The correlation factor $c$ is first fetched from memory. If $c$ is lower than a threshold $T$, then the neuron is evaluated in base precision. Otherwise, the binarized dot product result, $p_{bin}$, is computed, and the fitted line is used to obtain the estimated base precision result $\hat{p}_{base} = m * p_{bin} + b$. If batch normalization and residual connections are used, then $\hat{p}_{base}$ is transformed by using the batch normalization parameters of the base precision neuron, and the residual input is added. If the resulting estimated ReLU input is negative, a zero ReLU output is predicted, skipping evaluation of this neuron. Otherwise, the neuron is evaluated in base precision.

Since some neurons have low self-correlation with their 1-bit counterparts, using 1-bit predictors for the entire network will incur in significant accuracy loss. Note that incorrectly predicting a ReLU output as zero will result in accuracy loss, as incorrect neuron outputs will be used, whereas incorrectly predicting an output as non-zero results represents a lost opportunity for saving computations but it has no impact on accuracy since in this case, the neuron is evaluated in base precision. To avoid accuracy loss, we leverage the aforementioned $T$ threshold and only apply our prediction scheme for neurons whose correlation is higher than $T$. We use the training data to set appropriate values for $T$ for each DNN, and verify its correctness using the unseen test data set. Note that $T$ can be used to control the trade-off between computation savings and accuracy: the higher the threshold the lower the accuracy loss but the smaller the savings.

\begin{figure}[t]
\centering
\includegraphics[width=0.8\columnwidth]{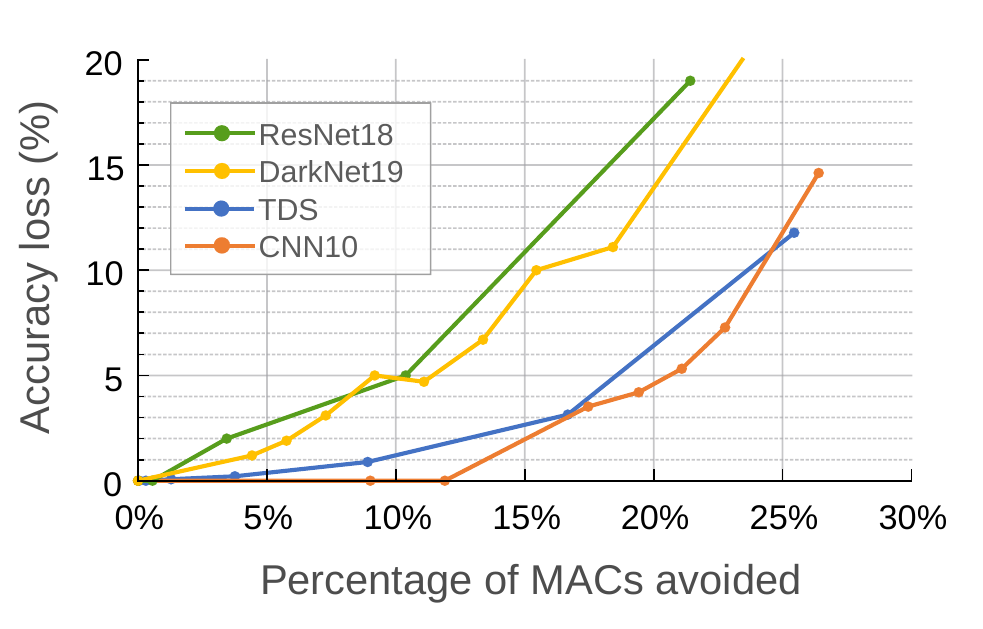}
\caption{Effect of the correlation-based threshold on accuracy loss and percentage of operations saved for different DNNs.}
\label{fig:wer_loss_correlation_threshold}
\end{figure}

Figure~\ref{fig:wer_loss_correlation_threshold} shows the effect that different thresholds have on the accuracy loss and percentage of operations saved for our set of DNNs. Each line corresponds to a different DNN, and each point is obtained by using a different threshold $T$ for linear correlation. The threshold is reduced from 1 (first point on the left for each line) to 0.6 (last point in the right). As it can be seen, the correlation threshold has a high impact in accuracy and percentage of savings. Furthermore, despite all the efforts to avoid incorrect predictions, the binarized predictor provides modest savings, 12\% of computations for CNN10 and much less for the other networks, if accuracy loss is maintained. Lower thresholds result in larger savings, but at the cost of introducing a significant amount of errors, as the correlation between binarized and base precision neuron is lower. The conclusion of this study is that the binary predictor alone can provide very low benefits, and this motivates our proposal for a hybrid predictor.

Previous work proposed to use several bits~\cite{cao2019seernet}, i.e. 4-bits, to improve self-correlation. However, we argue that 4-bits results in a significant overhead, and we propose in the next subsection an alternative solution that exploits spatial correlation to avoid mistakes done by the binarized predictor while incurring negligible overhead.

\subsubsection{Exploiting Spatial Correlation}\label{sec:angle_predictor}

Our \textit{Mixture-of-Rookies} predictor includes another scheme that exploits correlation among neurons with the same input vectors. This scheme aims to identify groups of neurons that share the same input vector and whose outputs are either all zero or all non-zero. In this manner, only one representative neuron of each group is evaluated during inference and, if it produces a zero output, the rest of the neurons of the group are assumed to produce a zero output without evaluating them. If the representative neuron produces a non-zero output, all neurons in the group are evaluated normally. 

The key challenge is to identify a minimum set of groups with high zero/non-zero correlation among them. To this end, we analyze the relation between the angle of any two vectors to model the probability that the dot product between both of these vectors and a given third vector will result on values with the same sign. Since the sign of the dot product depends only on the angle between the operand vectors, we can assume a distribution for the third vector and model the probability of having same-sign results as a relation between said angle.

Given two vectors $A$ and $B$, the dot-product between them is expressed as:
\begin{equation}
    A\cdot B = |A|\times|B|\cos{\theta}
\end{equation}
Where $\theta$ is the angle between $A$ and $B$. 

Since $\vert A\vert$ and $\vert B\vert$ are positive quantities, the sign of the result is given by the sign of $\cos{\theta}$, so it is entirely determined by $\theta$. For convenience, we can limit the study to angles in the range $[0, 180]$ ($\theta$ is the small angle between the vectors) and conclude that the output will be negative only when $\theta$ is below $90^\circ$.
\begin{equation}
sign(\cos{\theta}) =
\left\{
	\begin{array}{ll}
		+  & \mbox{if } \theta < 90^\circ \\
		-  & \mbox{if } \theta > 90^\circ
	\end{array}
\right.
\end{equation}

Note that the dot-product between 2 perpendicular vectors ($\theta = 90^\circ$) is $0$, so its sign can be defined however is most convenient.

\begin{figure}[t]
\centering
\includegraphics[width=0.8\columnwidth]{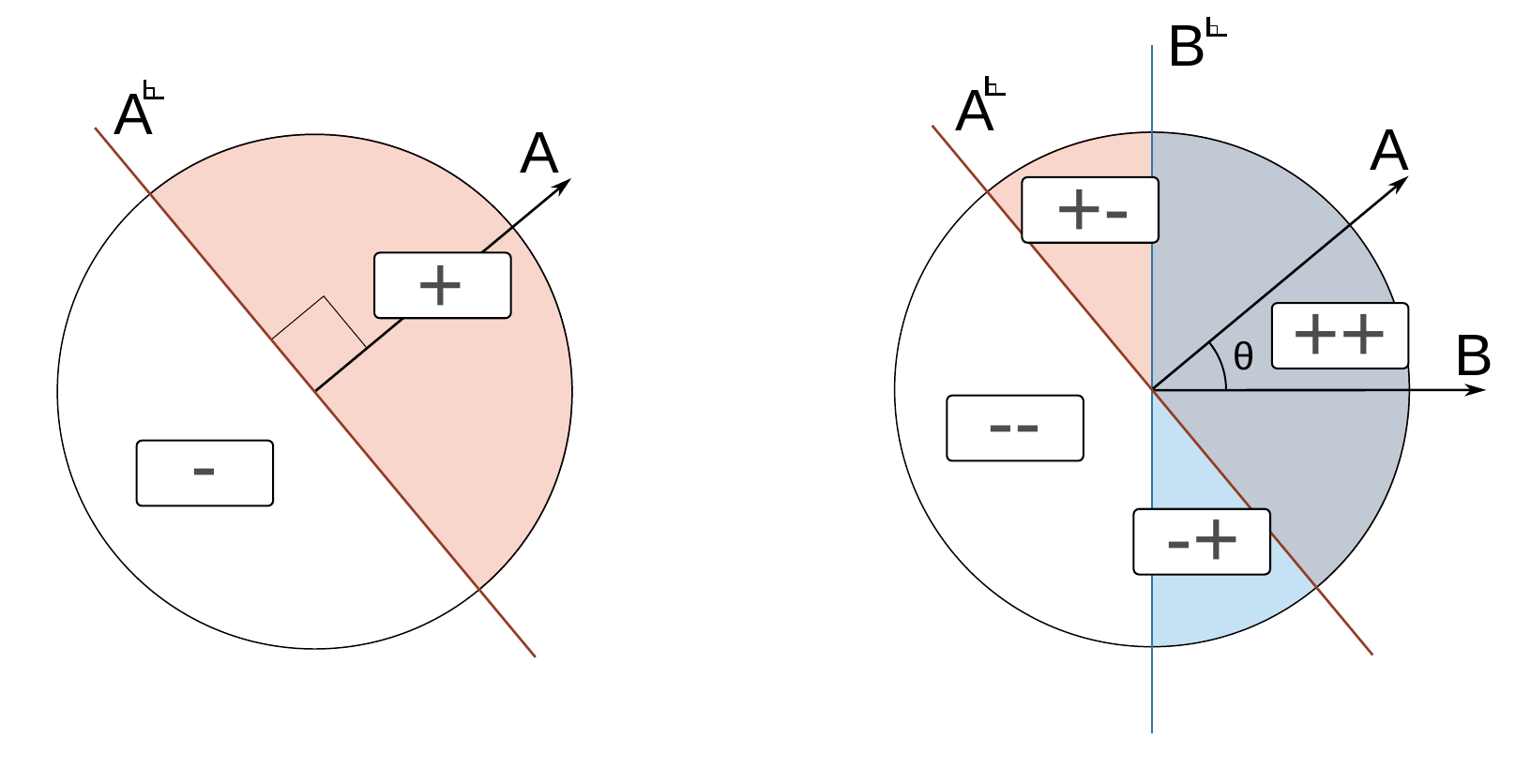}
\caption{The line perpendicular to $A$ partitions the circle into $2$ sectors. Given a random $C$, the sign of $C\cdot A$ is determined by the partition in which $C$ falls. If another vector $B$ is added (right figure), the circle is partitioned into 4 sectors, which determine the signs of $C\cdot A$ and $C\cdot B$.}
\label{fig:angle_thing}
\end{figure}

Figure \ref{fig:angle_thing} represents a circle and a vector $A$. If we draw a line perpendicular to $A$, the circle is divided in two halves. The dot-product between $A$ and any vector from the half in which $A$ is contained will result in a positive number. Correspondingly, the dot-product between $A$ and any vector from the other half of the circle will result in a negative number. If we add a second vector $B$ and its corresponding perpendicular line, the 4 regions (namely $R^{++}, R^{--}, R^{+-}, R^{-+}$) obtained by the overlapping of the 2 halves given by each vector characterize the range of vectors whose dot-products with $A$ and $B$ will result in each possible combination of signs $(++, --, +-, -+)$, and thus, it defines the probability of each possible outcome from $sign(C\cdot A)$ and $sign(C\cdot B)$ for a random vector $C$ as the probability of $C$ belonging to each of the previously defined regions. 

Assuming that $C$ follows a uniform distribution in the space (modeled as a hyper-sphere), these probabilities are given by the following expressions:
\begin{align}
    p(C \in R^{+-} | \theta) &= \frac{\theta}{360} \\
    p(C \in R^{-+} | \theta) &= \frac{\theta}{360} \label{eq:p_error}\\
    p(C \in R^{++} | \theta) &= \frac{1}{2} - \frac{\theta}{360} \\
    p(C \in R^{--} | \theta) &= \frac{1}{2} - \frac{\theta}{360}
\end{align}
Where $\theta$ is the angle between the vectors $A$ and $B$ expressed as a degree magnitude between $0^\circ$ and $180^\circ$.

When this circle is expanded to a sphere, the area relation between the regions (and hence the probabilities defined above) are preserved as a volume relation. We verified that this analysis holds for higher dimensions through a \textit{Montecarlo} simulation.

If we use the dot-product between $A$ (corresponding to the input weights of a neuron) and a random vector $C$ (corresponding to an input vector) to predict the sign of the dot-product between $B$ (corresponding to the weights of another neuron that uses the same input vector) and $C$, the worst case is when $C$ is from the $-+$ region, because $A \cdot C$ will be negative, and thus $B \cdot C$ will be assumed to be negative, when in reality, $B \cdot C$ is positive. Since negative dot products result in a zero output when the ReLU activation function is applied, the output of the neuron with weights $B$ will be wrongly assumed to be zero, without evaluating it (we call this scenario a false positive). The probability of a random $C$ vector to be in the $-+$ region is given by expression~\ref{eq:p_error}.

As we can see, the probability of causing a false positive is $0$ if the weights of the neurons are parallel, and increase up to $50\%$ for perpendicular neurons.

\begin{figure}[t]
\centering
\includegraphics[width=0.8\columnwidth]{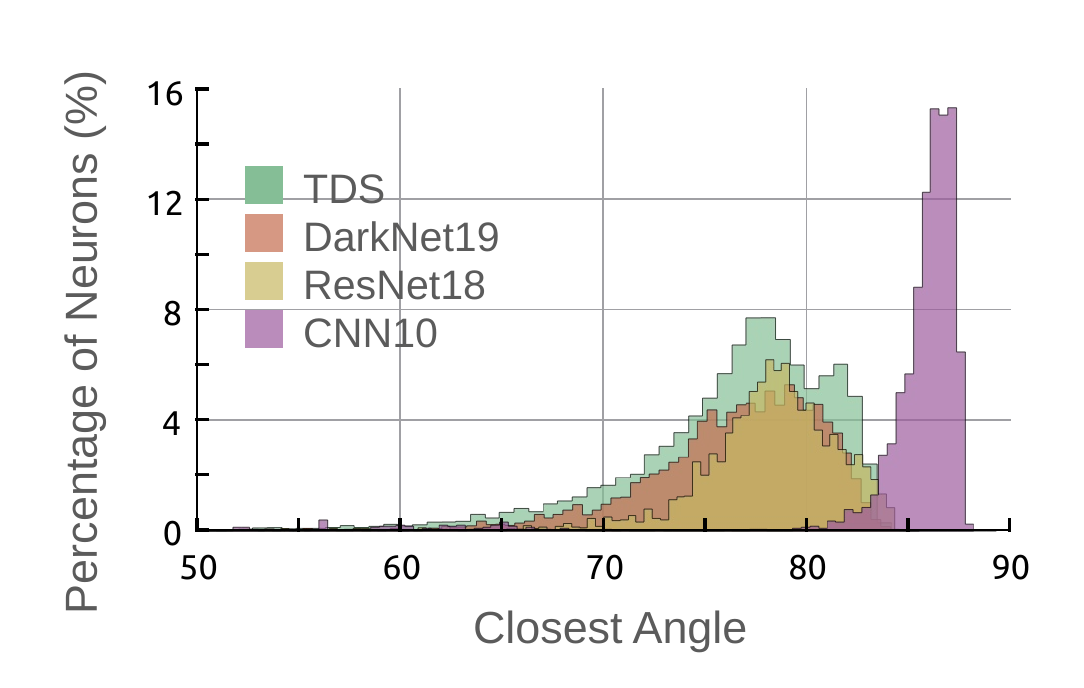}
\caption{Distribution of angles between each neuron and its closest neuron.}
\label{fig:angle_distribution}
\end{figure}

Since neurons' weights are represented as very high dimensional vectors, if they were random vectors, we would expect them all to be almost perpendicular, meaning that if there are two neurons within a layer with a $\theta$ lower than $90^\circ$, there is certain degree of correlation among them. To measure the amount of spatial correlation in the TDS layers, we computed, for each layer, all the neuron-neuron angles and then, for each neuron, obtained the angle with its closest neuron (the neuron with which is has the smaller angle). Figure~\ref{fig:angle_distribution} shows the distribution of such angles. If there was no correlation between neurons, we would expect most of them to fall between $80^\circ$ and $90^\circ$. However, as we can see, the majority of angles fall between $70^\circ$ and $80^\circ$, and a significant number of them are even lower.

Based on the previous observations, we propose a negative ReLU input predictor that leverages the correlation existing between pairs of neurons separated by an angle lower than $90^\circ$. 

To leverage this property, we could cluster each neuron with its closest neuron. However, an algorithm that directly applies this clustering strategy will likely create problematic arrangements such as chains of associated neurons that will end up in the same cluster, but with neurons that are very far apart. Instead, we propose an algorithm that generates clusters of neurons around a principal neuron, which we call \textit{proxy}, that will act as predictor for the rest of the cluster members. This algorithm first generates a directed graph with the neurons as nodes, and edges linking each neuron with its closest neuron. Then, the nodes are sorted by descending order of \textit{indegree} (number of incident edges) and, starting from the node with higher indegree, the node is removed from the graph and included in the set of proxies, whereas all the nodes linked to it are removed, too, and included as members of the previous node's cluster. This process is iterated until there are no more nodes in the graph.

By looking at the distribution of closest angles among neurons (figure~\ref{fig:angle_distribution}), it is clear that this technique alone will not result in good prediction accuracy. However, it provides useful information that can be leveraged to improve the performance of the self-correlation predictor. We combine this predictor with the self-correlation predictor described in the Section~\ref{sec:binary_predictor}. This predictor incurs in negligible overhead since it only requires the neurons to be arranged in a specific way in memory (including an index value to re-arrange the outputs) and minor additional control logic.

\begin{figure}[t]
\centering
\includegraphics[width=0.8\columnwidth]{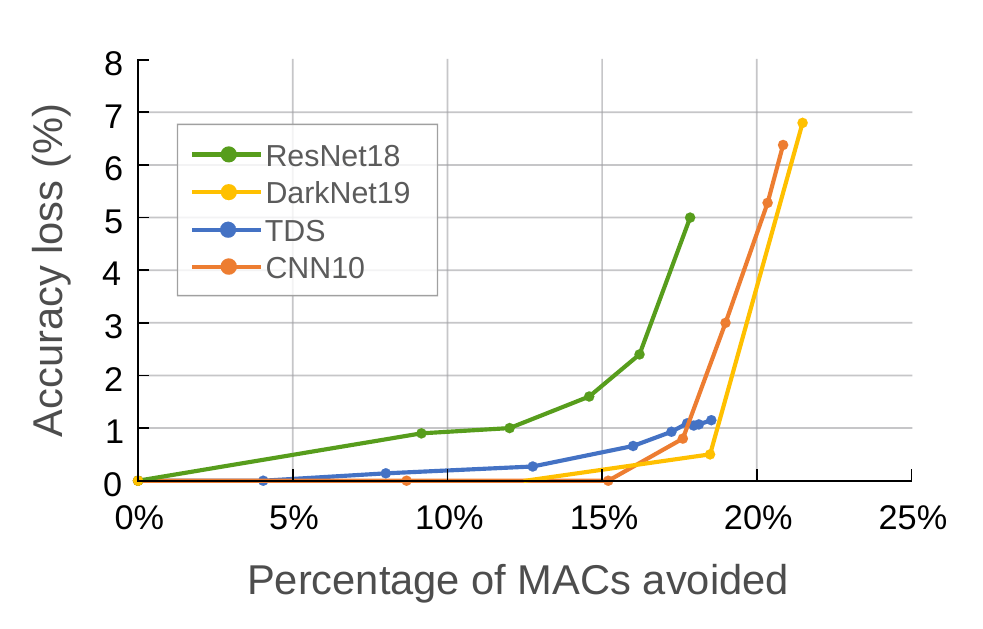}
\caption{Accuracy loss versus percentage of computations avoided for the hybrid \textit{Mixture-of-Rookies} predictor.}
\label{fig:wer_loss_hybrid_threshold}
\end{figure}

Figure~\ref{fig:wer_loss_hybrid_threshold} shows how adding spatial correlation information improves the results of the ReLU output predictor. Compared to the binarized predictor in isolation, whose results are shown in Figure~\ref{fig:wer_loss_correlation_threshold}, the predictor that employs both self-correlation and spatial correlation achieves larger computation savings with small accuracy loss.

\section{DNN Accelerator with ReLU Output Predictor} \label{sec:accelerator}

In this section we present a DNN accelerator that leverages our \textit{Mixture-of-Rookies} predictor, described in Section~\ref{sec:binary_predictor}, for energy-efficient DNN inference. Our accelerator is designed targeting use cases for inference in low-power devices, to support applications such as image or speech recognition on-edge. Hence, it is key to use very low area and power. Another constraining assumption for these use cases is that the input will be processed frame-by-frame (or image-by-image in the case of image recognition applications). For example, in~\cite{pratap2020scaling}, the authors explore very small dependency windows for the outputs of the TDS network presented in~\cite{hannun2019sequence} in order to minimize word-to-transcription latency, which is desirable for many applications of speech recognition on-edge.

\begin{figure}[t]
\centering
\includegraphics[width=0.8\linewidth]{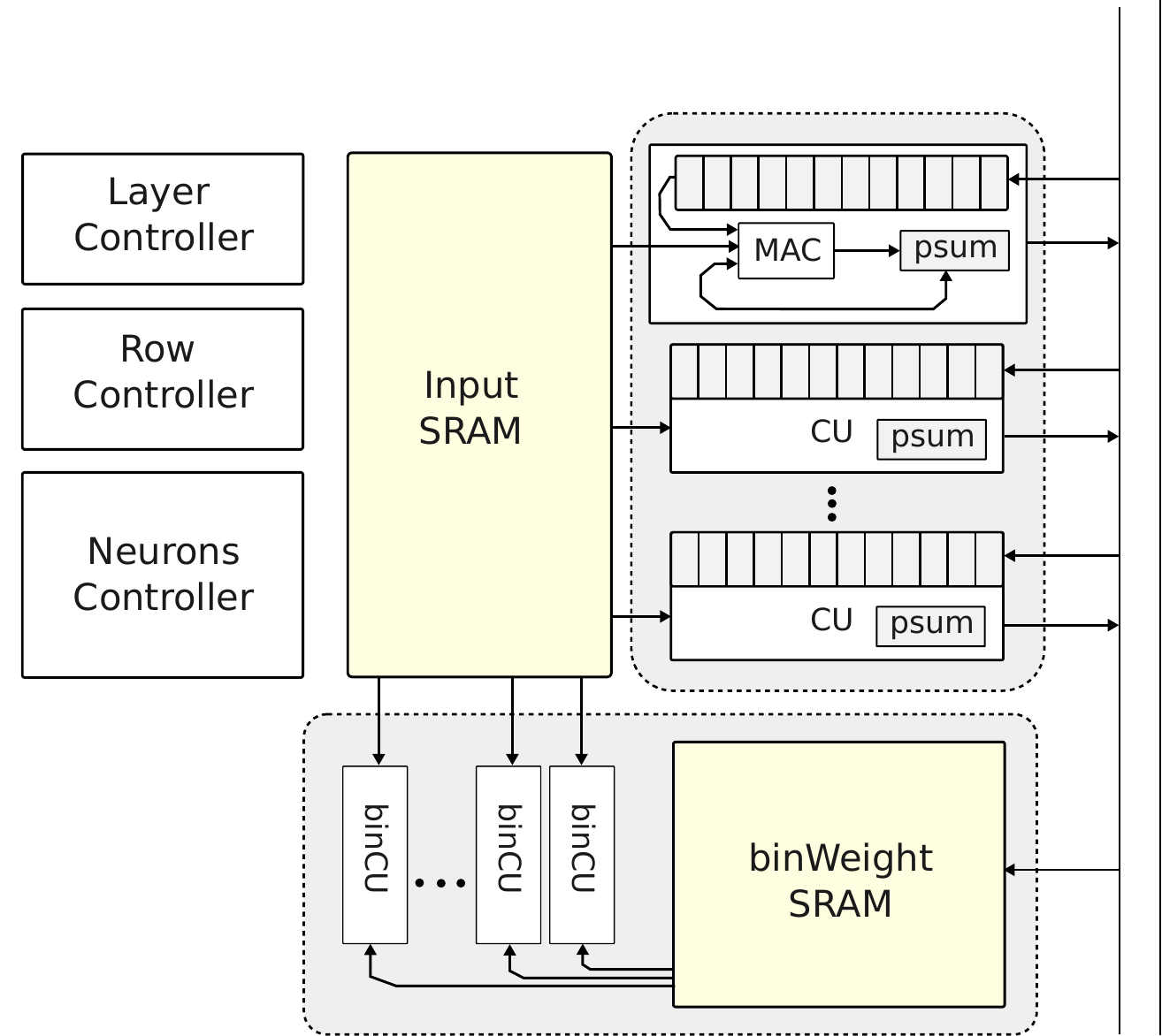}
\caption{Accelerator with support for \textit{Mixture-of-Rookies} ReLU output predictor.}
\label{fig:accel}
\end{figure}

Figure~\ref{fig:accel} illustrates the architecture of the accelerator. It contains three main control units: a \textit{Layer Controller}, a \textit{Row Controller} and a \textit{Neurons Controller}, an SRAM memory to store the \textit{inputs}, a set of \textit{Compute Units (CUs)} to compute neurons and a \textit{binary predictor} composed of an SRAM memory to store the \textit{binary weights} and a set of \textit{binary CUs} to compute the binarized neurons.

Each CU is responsible for the computations of the neurons assigned to it. The design has a configurable number of CUs, each with an interface to external memory. When a neuron is assigned to a CU, it generates requests to external memory and performs the required computation, accumulating the partial results on an internal register. To boost computations, the CUs perform in parallel several multiplications belonging to the same output. The number of parallel multipliers per CU is another parameter of the design.

\subsection{Control Unit}
The DNN processing is triggered by issuing an external request to the \textit{Layer Controller}, which generates \textit{Row Controller} requests to evaluate each layer of the DNN consecutively. The \textit{Row Controller} divides the output of the layer in rows, and issues memory requests to load the required inputs to compute them. Since an output row will generally require many inputs, and consequently, a large input SRAM, the inputs for the row are divided in blocks, which are loaded sequentially. This allows us to keep the input SRAM small. To leverage inputs reuse in CNN shift-windows, the inputs are loaded taking CNN \textit{stride} into account. Once an input window is loaded, the row controller issues a request to the \textit{Neuron Controller}, which generates issues to the CUs and binCUs to compute the neurons for that input window.

Since our predictor creates dependencies between proxy neurons and their cluster members, we have to compute first the proxies to unlock the corresponding non-proxy neurons (see Section~\ref{sec:angle_predictor}). Conceptually, the idea is to evaluate first all the proxies and generate a mask of neurons that are predicted to have a ReLU output of zero. Next, we evaluate the second predictor, binary predictor, for these neurons with predicted zero output and update the mask to include only those that are also predicted to produce a zero with the binary predictor. At this point, all the neurons not predicted to have zero ReLU output are assigned to CUs and their results are written back into memory. 

The evaluation of the binary predictor can be overlapped with the evaluation of the proxies. As soon as a neuron is predicted to have zero ReLU output by the corresponding proxy, the neurons controller issues requests to the binary CUs to compute the prediction based on the binarized neuron. If binary predictor also indicates a zero output, the output of that neuron is predicted to be zero, and a $0$ is written to external memory. Otherwise, the neuron is assigned to a free CU for full precision computation. 

The computation of the proxy neurons does not generates any overhead in execution time since they would have to be computed anyway, and neither does the evaluation of the self-correlation binarized predictor, which is performed in parallel with the rest of the neurons.

In our hardware implementation, we do not store the entire mask in memory as we interleave the evaluation of proxies and non-proxy neurons. As soon a proxy is evaluated, the corresponding non-proxy neurons are assigned higher priority than proxies, meaning that as long as there are available non-proxy neurons, they will be assigned to any free CU. Only when there is none, the proxies are assigned to CUs. Note that we still require a small buffer to keep track of the available cluster members. However, this implementation provides two advantages: a) the buffer is smaller than the memory required to store the mask and b) it does not impose a maximum output size.

\subsection{DNN Format}
\begin{figure}[t]
\centering
\includegraphics[width=0.8\linewidth]{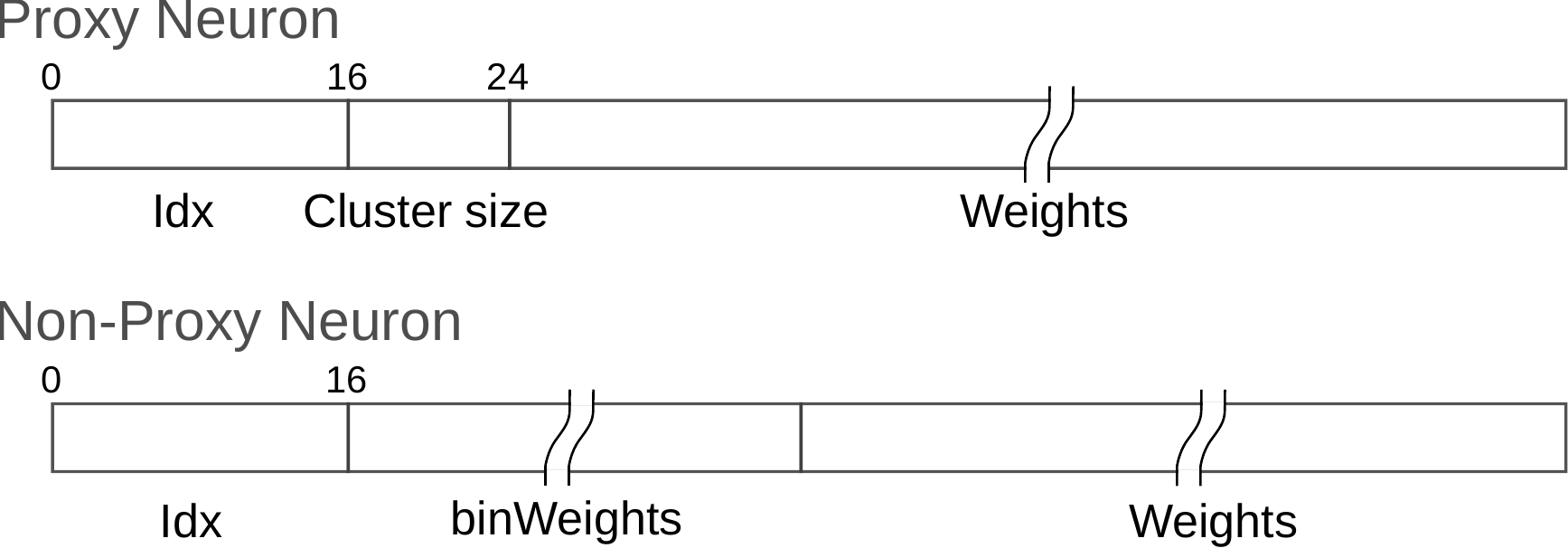}
\caption{Format to store the DNN in external memory.}
\label{fig:dnn_format}
\end{figure}

In order to support the execution flow previously described, we provide format to store the DNN in main memory, illustrated in Figure~\ref{fig:dnn_format}. The DNN layers are divided in two tables. The first table contains the proxy neurons. Each row contains an \textit{index (idx)} field to indicate the original position of the neuron, a \textit{cluster size} field to indicate the number of neurons in its cluster and finally, the weights of the neuron. The second table contains the non-proxy neurons sorted by the cluster they belong to. This means that the neurons associated with the first proxy in the proxy table occupy the first positions, the next positions are filled with the neurons associated with the second proxy and so on. Each row in this table contains the weights, the binary weights and an index to their original position. Since the binary weights are the sign bits from the weights, the sign bit is removed from every weight to offset the memory footprint overhead of storing the binary weights, and to avoid any increase in memory reads, which would otherwise affect the performance benefits.

\subsection{Compute Units}
The accelerator contains a group of CUs to evaluate neurons. Each CU operates independently and is connected to main memory through its own ports. They are assigned neurons by the neurons controller and are responsible for the computation of that neuron. As soon as a CU receives a request to process a new neuron, it starts fetching its weights from external memory to the internal buffer. Next, it reads inputs sequentially from the input SRAM and performs the dot product between inputs and weights, whose result is temporarily stored in its \textit{partial sum (psum)} register until the dot product is completed. The CUs have a design parameter to adjust the MAC unit width, which allows for computing several MACs in parallel. Note that the port width of the weight buffer is adjusted correspondingly.

\subsection{Binary Prediction Unit}
To support our prediction scheme, we include a \textit{Binary Prediction Unit}. It is composed of an SRAM memory to store the binary weights for the non-proxy neurons and a set of binCUs. The binCUs are similar to the CUs, they process neurons and operate independently. However, they do not contain a weight buffer because they do not have to access external memory. Instead, they read binary inputs and binary weights from input memory and the binary weight SRAM, respectively, to perform the binary dot-product. Since these units perform binary multiplications and counting instead of a full MAC, their circuit is much simpler than the \textit{CUs}~\cite{silfa2019neuron}.

\section{Evaluation Methodology}

In order to assess the performance of our ReLU output predictor, we simulate the execution of four DNNs on the proposed accelerator presented in Section~\ref{sec:accelerator}. In this section, we describe the DNNs, tools and data sets employed for evaluating our proposal.

\subsection{DNNs}
We test our technique on four DNNs for two different applications. The first application is  low-latency \textit{Speech Recognition}, we use the TDS network described in Pratap et al.~\cite{pratap2020scaling} trained on \textit{Librispeech}~\cite{panayotov2015librispeech}. The second application is object recognition. We employ \textit{DarkNet19}~\cite{redmon2017yolo9000} and \textit{ResNet18} trained on \textit{ImageNet}~\cite{he2016deep}, and a CNN trained on \textit{CIFAR-10}~\cite{krizhevsky2009learning}.

We use a pre-trained version of the \textit{TDS} network from \textit{Facebook's} public repository. The TDS network was trained for $110$ epochs on the complete 1k-hour \textit{Librispeech} train set, in addition to pseudo-labeled data from \textit{LibriVox} as described in~\cite{synnaeve2019end}. The \textit{TDS} network is part of an \textit{End-To-End} speech recognition system. It receives pre-processed frames from an audio signal that encodes an \textit{utterance} and generates acoustic probabilities over a set of pre-defined \textit{Word-Pieces} which are consumed by a decoder. The decoder combines the acoustic probabilities with language-level probabilities given by a language model and generates a transcription.

The accuracy of a speech recognition system is measured in \textit{Word Error Rate} (\textit{WER}), which is computed by taking the minimum number of additions, substitutions and deletions required to transform the generated transcription to the reference transcription (\textit{edit distance}) and dividing it by the number of words in the reference transcription. For the experiments, we use the \textit{test\_other} data set from \textit{Librispeech}, that includes utterances with noise that are challenging to transcribe. TDS achieves a WER of 8.24\% in \textit{test\_other}.

Regarding the three CNNs for object recognition, we trained \textit{ResNet18} on ImageNet training data set, achieving top-1 accuracy of 70.7\%. We also trained \textit{Darknet19}~\cite{Darknet19} on ImageNet, achieving top-1 accuracy of 73.8\%. Darknet19 consists of 19 convolutional layers following the architecture shown in Figure~\ref{f:block_diagram_conv}. Finally, we trained a CNN for CIFAR-10 task, which we call \textit{CNN10}. This network consists of ten convolutional layers and achieves top-1 accuracy of 75.1\%.

\subsection{Models}
In order to obtain execution time, we developed a simulator that accurately models the architecture of the accelerator described in Section~\ref{sec:accelerator}. This simulator integrates \textit{DRAMsim3}~\cite{li2020dramsim3} to model main memory. We also implemented in \textit{Verilog} the main components of the accelerator, that were synthesized with \textit{Design Compiler} to obtain area and static power estimates. In order to estimate dynamic power, the modules implemented in Verilog were simulated with Synopsys VCS with random inputs to generate toggle rate files which we fed to \textit{Power Compiler}. Area and power (static and dynamic) for the internal SRAM memories were obtained from the custom version of \textit{CACTI} included in \textit{McPat}~\cite{li2009mcpat}.

\subsection{Parameters}
Table~\ref{tbl:parameters} lists the parameters used for the simulated system. Our accelerator runs at the same frequency as the external memory. Both accelerators (with and without predictor) are exactly the same, except for the \textit{binWeight SRAM} and the \textit{binCUs}, which are only used by the predictor. The number and width of \textit{CUs} was set to 8 for a maximum throughput of 64 MACs/cycle. We employ a baseline precision of 8 bits per each weight and input. We model 1 GB of \textit{LPDDR4 SDRAM} as main memory.

\begin{table}[t]
    \caption{Simulation parameters.} 
    \label{tbl:parameters}
    \centering
    \begin{tabular}{rl}
        \multicolumn{2}{c}{\textbf{DNN Accelerator}}\\
        \rowcolor{lightgray} Frequency & 1200 MHz\\
        \rowcolor{white} Input SRAM & 16 KB \\
        \rowcolor{lightgray} BinWeight SRAM & 2 KB \\
        Number binCUs & 8 \\
        \rowcolor{lightgray} Number of CUs & 8 \\
        CU width & 8 \\
        \rowcolor{lightgray} CU precision & 8 b \\
        CU Buffer & 1 KB \\
        \rowcolor{lightgray} binCU buffer & 0.56KB \\
     
        \multicolumn{2}{c}{\textbf{External LPDDR4 Memory}}\\
        \rowcolor{lightgray} Frequency & 1200 MHz\\
        Capacity & 1 GB \\
        \rowcolor{lightgray} Port Width & 8 B \\
        Burst Size & 64 B \\
    \end{tabular}
    
\end{table}
\section{Results}




In this section, we present the speedups and energy savings achieved by our hybrid ReLU output predictor, introduced in Section~\ref{sec:technique}, when implemented on top of a DNN accelerator as described in Section~\ref{sec:accelerator}. We compare our proposal with a baseline accelerator that does not include the \textit{binWeights} SRAM and the \textit{binCU} units, so the energy consumed by them is considered and reported as overhead for our predictor.

\begin{figure}[t]
\centering
\includegraphics[width=0.8\columnwidth]{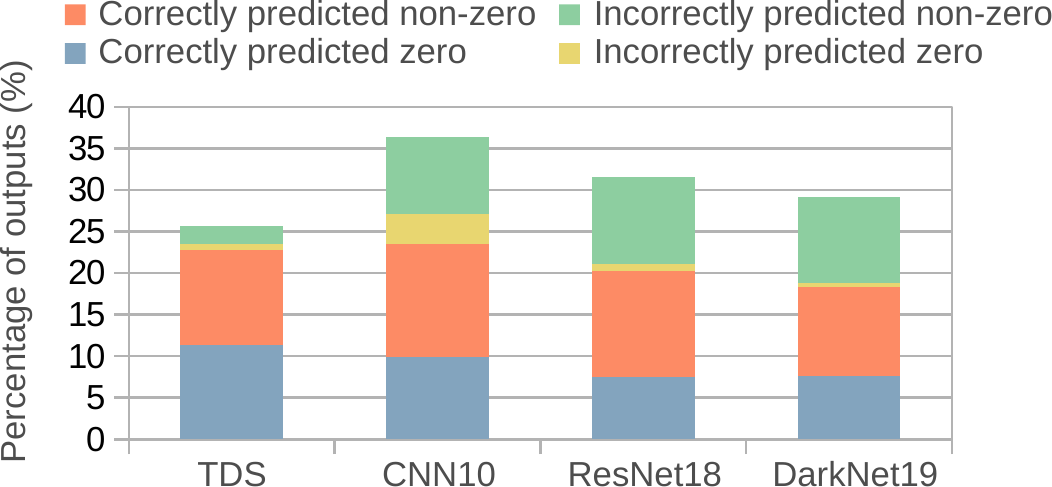}
\caption{Percentage of outputs that are correctly and incorrectly predicted as zero or nonzero by our \textit{Mixture-of-Rookies} predictor.}
\label{fig:predictor_accuracy}
\end{figure}

We first evaluate the accuracy of the proposed \textit{Mixture-of-Rookies} predictor. Figure~\ref{fig:predictor_accuracy} shows the percentage of correct and incorrect predictions. ``Correctly predicted zero'' means that the predictor indicates that the ReLU output will be zero and it is correct. In this case, neuron evaluation is skipped, avoiding all the related computations and memory accesses for base precision neuron, without affecting DNN classification accuracy. This is the case for 7\%-11\% of the outputs in our DNNs. ``Incorrectly predicted zero'' means that our scheme predicts a zero output, but the base precision output is non-zero. Neuron evaluation is also avoided, but these mispredictions may impact classification accuracy as they introduce errors in the DNN, since the output of these neurons is incorrectly set to zero. As it can be seen in Figure~\ref{fig:predictor_accuracy}, this type of mispredictions are fairly infrequent: 0.65\%, 0.8\%, 0.4\% and 3.6\% for TDS, Resnet18, Darkent19 and CNN10 respectively. We verified that the impact on DNN accuracy due to these mispredictions is lower than 1\% in our DNNs.

On the other hand, ``incorrectly predicted nonzero'' shows neurons where a non-zero output is predicted but the ReLU output is zero. These mispredictions have no impact on accuracy, as the neuron is evaluated when \textit{Mixture-of-Rookies} predicts non-zero, but they represent a missed opportunity for saving computations. Finally, ``correctly predicted nonzero'' category shows that between 10\% and 13\% of the outputs are non-zero values correctly identified by our predictor. Note that the four categories shown in Figure~\ref{fig:predictor_accuracy} do not add 100\% as there are neurons where the predictor is not applied for several reasons. First, our scheme is not applied in DNN layers that do not use ReLU activation function, this is common in TDS network. Second, \textit{proxy} neurons, i.e. centroids, are always evaluated in our scheme. Third, our predictor is disabled for neurons that show poor linear correlation with their binarized versions.

\begin{figure}[t]
     \centering
     \subfloat[Speedup.]{
        \label{fig:results_time}
        \includegraphics[width=0.8\columnwidth]{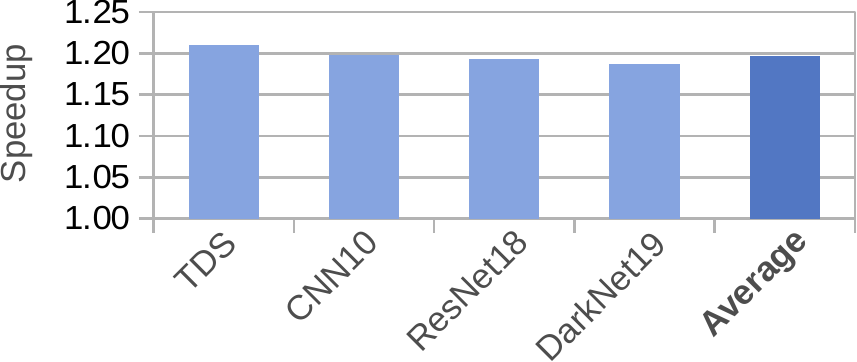}
    }
    
    \subfloat[Energy savings.]{
        \label{fig:results_energy}
        \includegraphics[width=0.8\columnwidth]{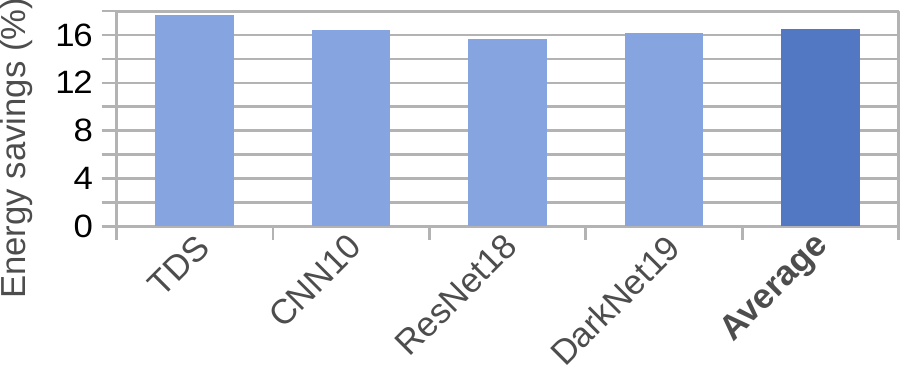}
    }
    \caption{Performance and energy savings achieved by our \textit{Mixture-of-Rookies} ReLU output predictor compared to the baseline.}
    \label{f:results}
\end{figure}

Figure~\ref{fig:results_time} shows the speedups achieved by our \textit{Mixture-of-Rookies} predictor. Our prediction scheme provides consistent and significant performance improvements, providing 19.8\% speedup on average. These speedups are due to skipping neuron evaluation when the predictor indicates that its ReLU output will be zero. Note that this avoids both computation and memory accesses, since weights for those neurons do not have to be fetched from main memory. Computations related to the predictor itself are largely overlapped with useful computations and, hence, they do not introduce any performance penalty.

Regarding energy consumption, Figure~\ref{fig:results_energy} shows the energy savings achieved by our ReLU output predictor. As it can be seen, our technique obtains significant energy savings across all the applications, reducing energy by 16.5\% on average. Our system improves energy consumption because the number of computations and memory accesses is reduced proportionally with the number of neurons skipped. Furthermore, the hardware for our \textit{Mixture-of-Rookies} predictor represents a small overhead: 5.3\% in area and less than 1\% in energy consumption (included in the results). These overheads are more than offset by the benefits so the net result is an important reduction in energy consumption.

\section{related work}

Predicting the output of instructions to save time and energy can be traced back to pioneer work performed by Lipasti et al.~\cite{lipasti1996value}, Gabbay et al.~\cite{gabbay1996speculative} and Gonzalez et al.~\cite{gonzalez1997speculative}, where \textit{Value Locality} is defined and studied to complement the already widespread \textit{Spatial} and \textit{Temporal Memory Localities}. These papers were followed by a plethora of work \cite{lipasti1996exceeding, wang1997highly, sazeides1997implementations, roth1998dependence, marcuello1999clustered, calder1999selective, goeman2001differential} proposing techniques to exploit value predictability at different levels. Proposals to predict outputs within neural networks, such as this work, can be regarded as a continuation of that line of research. The main difference is that in the case of DNN prediction schemes, the computation unit upon which prediction is applied is no longer an instruction, but a MAC operation, a dot-product or a full neuron. Furthermore, the target in DNNs is to predict whether the output is zero or non-zero.

Exploiting weight sparsity is a popular approach to reduce computations and external memory accesses in modern \textit{DNNs}. Many works propose techniques based on generating weight sparsity by zeroing-out less relevant weights \cite{zhang2018systematic, iandola2016squeezenet, yu2017scalpel, ding2019centripetal, dai2019grow, wen2016learning, ma2020pconv, liu2020autocompress, deng2018permdnn}, which is known a \textit{pruning}. This sparsity is leveraged by using a compact DNN representation that does not include zeros, and \textit{sparse DNN} accelerators. 

Related to weight pruning is the idea of exploiting input sparsity \cite{albericio2016cnvlutin, judd2017cnvlutin2, deng2018permdnn, qin2020sigma, chen2016eyeriss}. This approach consists on generating the output in a compact sparse format, similar to the format used for pruned weights, which seamlessly allows a \textit{sparse DNN} accelerator~\cite{qin2020sigma, whatmough201714, zhang2016cambricon, gupta2019masr, parashar2017scnn, han2017ese} to skip more computations. Our approach is different because we focus on output sparsity. The challenge with output sparsity is predicting whether an output will be zero before computing it. The advantage is that once a zero output is identified, the whole neuron computation can be avoided, and since the \textit{ReLU} activation function generate huge output sparsity (around $88\%$ for the TDS network) there is great potential. Moreover, the three different approaches (weight pruning, input sparsity and output sparsity) are complementary so a DNN accelerator can benefit from all of them.

Closer to our approach are proposals to exploit output sparsity \cite{akhlaghi2018snapea, lin2017predictivenet, song2018prediction, cao2019seernet, dong2017more}. These works generally exploit self-correlation by performing first an approximated computation of the dot-product and depending on the result, decide whether the output will be negative (and thus, zero after ReLU) or have to be computed in full precision. Our approach includes a component to exploit self-correlation, but it is based on a much smaller and simpler 1-bit predictor refined to improve its accuracy, whereas these previous works include predictors with much higher overheads.

Another approach consists in exploiting \textit{Spatial correlation}~\cite{shomron2018spatial, shomron2020thanks, kim2018mosaic, figurnov2015perforatedcnns, mahmoud2018diffy, kligvasser2018xunit}. However, it has been mainly applied for convolutional neural networks (CNNs), where the neurons within the same layers are correlated since they use the same weights. These works design sophisticated approaches to exploit spatial correlation in CNNs to save computations. However, those techniques rely directly on the kind of spatial correlation observed in CNNs and thus cannot be applied to FC layer, which dominate the computations for many modern DNNs. Our approach is different because we propose a technique to exploit general spatial correlation, which can be applied to both CONV and FC layers with no overhead. Moreover, the core of our \textit{Mixture-of-Rookies} technique resides in the synergistic combination of both predictors in a hybrid scheme which provides better prediction accuracy than any of the individual components with negligible cost. Because of the very low overhead incurred, we can obtain better benefits than previously proposed techniques.
\section{conclusions}

In this paper, we show that a large percentage of neurons in modern DNNs use a ReLU activation function that very often produces a zero output. Based on this observation, we propose a neuron zero-output predictor that combines two inexpensive schemes.

The first exploits the high linear correlation between
binarized (1-bit) and full-precision (8-bit) dot products to identify neurons that will produce negative input values to their ReLu activation functions. The second scheme clusters together neurons that have the same input vector and tend to produce negative ReLU inputs at the same time. If a neuron from a given group produces a negative ReLU input, it is predicted that the rest of neurons in the same group will also generate a negative ReLU input, avoiding all the associated computation. Our solution implements this hybrid predictor on top of a  state-of-the-art DNN accelerator. The experimental results show that the combination of both predictors provides 1.2x speedup and 16.5\% energy savings on average for a set of diverse DNNs, while introducing a small area overhead of 5.3\%.
\section*{ACKNOWLEDGMENTS}
This work has been supported by the CoCoUnit ERC Advanced Grant of the EU’s Horizon 2020 program (grant No 833057), the Spanish State Research Agency under grant PID2020-113172RB-I00 (AEI/FEDER, EU), the ICREA Academia program and the Spanish MICINN Ministry under grant BES-2017-080605.  

\bibliographystyle{IEEEtran}
\bibliography{IEEEabrv, refs}

\begin{thebibliography}{10}
\providecommand{\url}[1]{#1}
\csname url@samestyle\endcsname
\providecommand{\newblock}{\relax}
\providecommand{\bibinfo}[2]{#2}
\providecommand{\BIBentrySTDinterwordspacing}{\spaceskip=0pt\relax}
\providecommand{\BIBentryALTinterwordstretchfactor}{4}
\providecommand{\BIBentryALTinterwordspacing}{\spaceskip=\fontdimen2\font plus
\BIBentryALTinterwordstretchfactor\fontdimen3\font minus
  \fontdimen4\font\relax}
\providecommand{\BIBforeignlanguage}[2]{{%
\expandafter\ifx\csname l@#1\endcsname\relax
\typeout{** WARNING: IEEEtran.bst: No hyphenation pattern has been}%
\typeout{** loaded for the language `#1'. Using the pattern for}%
\typeout{** the default language instead.}%
\else
\language=\csname l@#1\endcsname
\fi
#2}}
\providecommand{\BIBdecl}{\relax}
\BIBdecl

\bibitem{pratap2020scaling}
V.~Pratap, Q.~Xu, J.~Kahn, G.~Avidov, T.~Likhomanenko, A.~Hannun,
  V.~Liptchinsky, G.~Synnaeve, and R.~Collobert, ``Scaling up online speech
  recognition using convnets,'' \emph{arXiv preprint arXiv:2001.09727}, 2020.

\bibitem{he2016deep}
K.~He, X.~Zhang, S.~Ren, and J.~Sun, ``Deep residual learning for image
  recognition,'' in \emph{Proceedings of the IEEE conference on computer vision
  and pattern recognition}, 2016, pp. 770--778.

\bibitem{akhlaghi2018snapea}
V.~Akhlaghi, A.~Yazdanbakhsh, K.~Samadi, R.~K. Gupta, and H.~Esmaeilzadeh,
  ``Snapea: Predictive early activation for reducing computation in deep
  convolutional neural networks,'' in \emph{2018 ACM/IEEE 45th Annual
  International Symposium on Computer Architecture (ISCA)}.\hskip 1em plus
  0.5em minus 0.4em\relax IEEE, 2018, pp. 662--673.

\bibitem{ImageNetDeng}
J.~{Deng}, W.~{Dong}, R.~{Socher}, L.~{Li}, {Kai Li}, and {Li Fei-Fei},
  ``Imagenet: A large-scale hierarchical image database,'' in \emph{2009 IEEE
  Conference on Computer Vision and Pattern Recognition}, 2009, pp. 248--255.

\bibitem{MarcComputationReuse}
M.~{Riera}, J.~{Arnau}, and A.~{Gonzalez}, ``Computation reuse in dnns by
  exploiting input similarity,'' in \emph{2018 ACM/IEEE 45th Annual
  International Symposium on Computer Architecture (ISCA)}, 2018, pp. 57--68.

\bibitem{AlexNet}
\BIBentryALTinterwordspacing
A.~Krizhevsky, I.~Sutskever, and G.~E. Hinton, ``Imagenet classification with
  deep convolutional neural networks,'' \emph{Commun. ACM}, vol.~60, no.~6, p.
  84–90, May 2017. [Online]. Available:
  \url{https://doi-org.recursos.biblioteca.upc.edu/10.1145/3065386}
\BIBentrySTDinterwordspacing

\bibitem{lin2017predictivenet}
Y.~Lin, C.~Sakr, Y.~Kim, and N.~Shanbhag, ``Predictivenet: An energy-efficient
  convolutional neural network via zero prediction,'' in \emph{2017 IEEE
  international symposium on circuits and systems (ISCAS)}.\hskip 1em plus
  0.5em minus 0.4em\relax IEEE, 2017, pp. 1--4.

\bibitem{cao2019seernet}
S.~Cao, L.~Ma, W.~Xiao, C.~Zhang, Y.~Liu, L.~Zhang, L.~Nie, and Z.~Yang,
  ``Seernet: Predicting convolutional neural network feature-map sparsity
  through low-bit quantization,'' in \emph{Proceedings of the IEEE/CVF
  Conference on Computer Vision and Pattern Recognition}, 2019, pp.
  11\,216--11\,225.

\bibitem{silfa2019neuron}
F.~Silfa, G.~Dot, J.-M. Arnau, and A.~Gonz{\`a}lez, ``Neuron-level fuzzy
  memoization in rnns,'' in \emph{Proceedings of the 52nd Annual IEEE/ACM
  International Symposium on Microarchitecture}, 2019, pp. 782--793.

\bibitem{pmlr-v37-ioffe15}
\BIBentryALTinterwordspacing
S.~Ioffe and C.~Szegedy, ``Batch normalization: Accelerating deep network
  training by reducing internal covariate shift,'' in \emph{Proceedings of the
  32nd International Conference on Machine Learning}, ser. Proceedings of
  Machine Learning Research, F.~Bach and D.~Blei, Eds., vol.~37.\hskip 1em plus
  0.5em minus 0.4em\relax Lille, France: PMLR, 07--09 Jul 2015, pp. 448--456.
  [Online]. Available: \url{http://proceedings.mlr.press/v37/ioffe15.html}
\BIBentrySTDinterwordspacing

\bibitem{song2018prediction}
M.~Song, J.~Zhao, Y.~Hu, J.~Zhang, and T.~Li, ``Prediction based execution on
  deep neural networks,'' in \emph{2018 ACM/IEEE 45th Annual International
  Symposium on Computer Architecture (ISCA)}.\hskip 1em plus 0.5em minus
  0.4em\relax IEEE, 2018, pp. 752--763.

\bibitem{judd2016stripes}
P.~Judd, J.~Albericio, T.~Hetherington, T.~M. Aamodt, and A.~Moshovos,
  ``Stripes: Bit-serial deep neural network computing,'' in \emph{2016 49th
  Annual IEEE/ACM International Symposium on Microarchitecture (MICRO)}.\hskip
  1em plus 0.5em minus 0.4em\relax IEEE, 2016, pp. 1--12.

\bibitem{shomron2018spatial}
G.~Shomron and U.~Weiser, ``Spatial correlation and value prediction in
  convolutional neural networks,'' \emph{IEEE Computer Architecture Letters},
  vol.~18, no.~1, pp. 10--13, 2018.

\bibitem{shomron2020thanks}
G.~Shomron, R.~Banner, M.~Shkolnik, and U.~Weiser, ``Thanks for nothing:
  Predicting zero-valued activations with lightweight convolutional neural
  networks,'' in \emph{European Conference on Computer Vision}.\hskip 1em plus
  0.5em minus 0.4em\relax Springer, 2020, pp. 234--250.

\bibitem{hannun2019sequence}
A.~Hannun, A.~Lee, Q.~Xu, and R.~Collobert, ``Sequence-to-sequence speech
  recognition with time-depth separable convolutions,'' \emph{arXiv preprint
  arXiv:1904.02619}, 2019.

\bibitem{synnaeve2019end}
G.~Synnaeve, Q.~Xu, J.~Kahn, T.~Likhomanenko, E.~Grave, V.~Pratap, A.~Sriram,
  V.~Liptchinsky, and R.~Collobert, ``End-to-end asr: from supervised to
  semi-supervised learning with modern architectures,'' \emph{arXiv preprint
  arXiv:1911.08460}, 2019.

\bibitem{BinaryNet}
\BIBentryALTinterwordspacing
M.~Courbariaux and Y.~Bengio, ``Binarynet: Training deep neural networks with
  weights and activations constrained to +1 or -1,'' \emph{CoRR}, vol.
  abs/1602.02830, 2016. [Online]. Available:
  \url{http://arxiv.org/abs/1602.02830}
\BIBentrySTDinterwordspacing

\bibitem{anderson2017high}
A.~G. Anderson and C.~P. Berg, ``The high-dimensional geometry of binary neural
  networks,'' \emph{arXiv preprint arXiv:1705.07199}, 2017.

\bibitem{panayotov2015librispeech}
V.~Panayotov, G.~Chen, D.~Povey, and S.~Khudanpur, ``Librispeech: an asr corpus
  based on public domain audio books,'' in \emph{2015 IEEE international
  conference on acoustics, speech and signal processing (ICASSP)}.\hskip 1em
  plus 0.5em minus 0.4em\relax IEEE, 2015, pp. 5206--5210.

\bibitem{redmon2017yolo9000}
J.~Redmon and A.~Farhadi, ``Yolo9000: better, faster, stronger,'' in
  \emph{Proceedings of the IEEE conference on computer vision and pattern
  recognition}, 2017, pp. 7263--7271.

\bibitem{krizhevsky2009learning}
A.~Krizhevsky, G.~Hinton \emph{et~al.}, ``Learning multiple layers of features
  from tiny images,'' 2009.

\bibitem{Darknet19}
``{I}mage{N}et {C}lassification,''
  \url{https://pjreddie.com/darknet/imagenet/}, accessed: 2021-04-09.

\bibitem{li2020dramsim3}
S.~Li, Z.~Yang, D.~Reddy, A.~Srivastava, and B.~Jacob, ``Dramsim3: a
  cycle-accurate, thermal-capable dram simulator,'' \emph{IEEE Computer
  Architecture Letters}, vol.~19, no.~2, pp. 106--109, 2020.

\bibitem{li2009mcpat}
S.~Li, J.~H. Ahn, R.~D. Strong, J.~B. Brockman, D.~M. Tullsen, and N.~P.
  Jouppi, ``Mcpat: An integrated power, area, and timing modeling framework for
  multicore and manycore architectures,'' in \emph{Proceedings of the 42nd
  Annual IEEE/ACM International Symposium on Microarchitecture}, 2009, pp.
  469--480.

\bibitem{lipasti1996value}
M.~H. Lipasti, C.~B. Wilkerson, and J.~P. Shen, ``Value locality and load value
  prediction,'' in \emph{Proceedings of the seventh international conference on
  Architectural support for programming languages and operating systems}, 1996,
  pp. 138--147.

\bibitem{gabbay1996speculative}
F.~Gabbay and A.~Mendelson, \emph{Speculative execution based on value
  prediction}.\hskip 1em plus 0.5em minus 0.4em\relax Citeseer, 1996.

\bibitem{gonzalez1997speculative}
J.~Gonz{\'a}lez and A.~Gonz{\'a}lez, ``Speculative execution via address
  prediction and data prefetching,'' in \emph{Proceedings of the 11th
  international conference on Supercomputing}, 1997, pp. 196--203.

\bibitem{lipasti1996exceeding}
M.~H. Lipasti and J.~P. Shen, ``Exceeding the dataflow limit via value
  prediction,'' in \emph{Proceedings of the 29th Annual IEEE/ACM International
  Symposium on Microarchitecture. MICRO 29}.\hskip 1em plus 0.5em minus
  0.4em\relax IEEE, 1996, pp. 226--237.

\bibitem{wang1997highly}
K.~Wang and M.~Franklin, ``Highly accurate data value prediction using hybrid
  predictors,'' in \emph{Proceedings of 30th Annual International Symposium on
  Microarchitecture}.\hskip 1em plus 0.5em minus 0.4em\relax IEEE, 1997, pp.
  281--290.

\bibitem{sazeides1997implementations}
Y.~Sazeides and J.~E. Smith, ``Implementations of context based value
  predictors,'' Citeseer, Tech. Rep., 1997.

\bibitem{roth1998dependence}
A.~Roth, A.~Moshovos, and G.~S. Sohi, ``Dependence based prefetching for linked
  data structures,'' in \emph{Proceedings of the eighth international
  conference on Architectural support for programming languages and operating
  systems}, 1998, pp. 115--126.

\bibitem{marcuello1999clustered}
P.~Marcuello and A.~Gonzalez, ``Clustered speculative multithreaded
  processors,'' in \emph{Proceedings of the 13th International Conference on
  Supercomputing}, 1999, pp. 365--372.

\bibitem{calder1999selective}
B.~Calder, G.~Reinman, and D.~M. Tullsen, ``Selective value prediction,'' in
  \emph{Proceedings of the 26th annual international symposium on computer
  architecture}, 1999, pp. 64--74.

\bibitem{goeman2001differential}
B.~Goeman, H.~Vandierendonck, and K.~De~Bosschere, ``Differential fcm:
  Increasing value prediction accuracy by improving table usage efficiency,''
  in \emph{Proceedings HPCA Seventh International Symposium on High-Performance
  Computer Architecture}.\hskip 1em plus 0.5em minus 0.4em\relax IEEE, 2001,
  pp. 207--216.

\bibitem{zhang2018systematic}
T.~Zhang, S.~Ye, K.~Zhang, J.~Tang, W.~Wen, M.~Fardad, and Y.~Wang, ``A
  systematic dnn weight pruning framework using alternating direction method of
  multipliers,'' in \emph{Proceedings of the European Conference on Computer
  Vision (ECCV)}, 2018, pp. 184--199.

\bibitem{iandola2016squeezenet}
F.~N. Iandola, S.~Han, M.~W. Moskewicz, K.~Ashraf, W.~J. Dally, and K.~Keutzer,
  ``Squeezenet: Alexnet-level accuracy with 50x fewer parameters and< 0.5 mb
  model size,'' \emph{arXiv preprint arXiv:1602.07360}, 2016.

\bibitem{yu2017scalpel}
J.~Yu, A.~Lukefahr, D.~Palframan, G.~Dasika, R.~Das, and S.~Mahlke, ``Scalpel:
  Customizing dnn pruning to the underlying hardware parallelism,'' \emph{ACM
  SIGARCH Computer Architecture News}, vol.~45, no.~2, pp. 548--560, 2017.

\bibitem{ding2019centripetal}
X.~Ding, G.~Ding, Y.~Guo, and J.~Han, ``Centripetal sgd for pruning very deep
  convolutional networks with complicated structure,'' in \emph{Proceedings of
  the IEEE/CVF Conference on Computer Vision and Pattern Recognition}, 2019,
  pp. 4943--4953.

\bibitem{dai2019grow}
X.~Dai, H.~Yin, and N.~K. Jha, ``Grow and prune compact, fast, and accurate
  lstms,'' \emph{IEEE Transactions on Computers}, vol.~69, no.~3, pp. 441--452,
  2019.

\bibitem{wen2016learning}
W.~Wen, C.~Wu, Y.~Wang, Y.~Chen, and H.~Li, ``Learning structured sparsity in
  deep neural networks,'' in \emph{Proceedings of the 30th International
  Conference on Neural Information Processing Systems}, 2016, pp. 2082--2090.

\bibitem{ma2020pconv}
X.~Ma, F.-M. Guo, W.~Niu, X.~Lin, J.~Tang, K.~Ma, B.~Ren, and Y.~Wang, ``Pconv:
  The missing but desirable sparsity in dnn weight pruning for real-time
  execution on mobile devices,'' in \emph{Proceedings of the AAAI Conference on
  Artificial Intelligence}, vol.~34, no.~04, 2020, pp. 5117--5124.

\bibitem{liu2020autocompress}
N.~Liu, X.~Ma, Z.~Xu, Y.~Wang, J.~Tang, and J.~Ye, ``Autocompress: An automatic
  dnn structured pruning framework for ultra-high compression rates,'' in
  \emph{Proceedings of the AAAI Conference on Artificial Intelligence},
  vol.~34, no.~04, 2020, pp. 4876--4883.

\bibitem{deng2018permdnn}
C.~Deng, S.~Liao, Y.~Xie, K.~K. Parhi, X.~Qian, and B.~Yuan, ``Permdnn:
  Efficient compressed dnn architecture with permuted diagonal matrices,'' in
  \emph{2018 51st Annual IEEE/ACM International Symposium on Microarchitecture
  (MICRO)}.\hskip 1em plus 0.5em minus 0.4em\relax IEEE, 2018, pp. 189--202.

\bibitem{albericio2016cnvlutin}
J.~Albericio, P.~Judd, T.~Hetherington, T.~Aamodt, N.~E. Jerger, and
  A.~Moshovos, ``Cnvlutin: Ineffectual-neuron-free deep neural network
  computing,'' \emph{ACM SIGARCH Computer Architecture News}, vol.~44, no.~3,
  pp. 1--13, 2016.

\bibitem{judd2017cnvlutin2}
P.~Judd, A.~Delmas, S.~Sharify, and A.~Moshovos, ``Cnvlutin2:
  Ineffectual-activation-and-weight-free deep neural network computing,''
  \emph{arXiv preprint arXiv:1705.00125}, 2017.

\bibitem{qin2020sigma}
E.~Qin, A.~Samajdar, H.~Kwon, V.~Nadella, S.~Srinivasan, D.~Das, B.~Kaul, and
  T.~Krishna, ``Sigma: A sparse and irregular gemm accelerator with flexible
  interconnects for dnn training,'' in \emph{2020 IEEE International Symposium
  on High Performance Computer Architecture (HPCA)}.\hskip 1em plus 0.5em minus
  0.4em\relax IEEE, 2020, pp. 58--70.

\bibitem{chen2016eyeriss}
Y.-H. Chen, T.~Krishna, J.~S. Emer, and V.~Sze, ``Eyeriss: An energy-efficient
  reconfigurable accelerator for deep convolutional neural networks,''
  \emph{IEEE journal of solid-state circuits}, vol.~52, no.~1, pp. 127--138,
  2016.

\bibitem{whatmough201714}
P.~N. Whatmough, S.~K. Lee, H.~Lee, S.~Rama, D.~Brooks, and G.-Y. Wei, ``14.3 a
  28nm soc with a 1.2 ghz 568nj/prediction sparse deep-neural-network engine
  with> 0.1 timing error rate tolerance for iot applications,'' in \emph{2017
  IEEE International Solid-State Circuits Conference (ISSCC)}.\hskip 1em plus
  0.5em minus 0.4em\relax IEEE, 2017, pp. 242--243.

\bibitem{zhang2016cambricon}
S.~Zhang, Z.~Du, L.~Zhang, H.~Lan, S.~Liu, L.~Li, Q.~Guo, T.~Chen, and Y.~Chen,
  ``Cambricon-x: An accelerator for sparse neural networks,'' in \emph{2016
  49th Annual IEEE/ACM International Symposium on Microarchitecture
  (MICRO)}.\hskip 1em plus 0.5em minus 0.4em\relax IEEE, 2016, pp. 1--12.

\bibitem{gupta2019masr}
U.~Gupta, B.~Reagen, L.~Pentecost, M.~Donato, T.~Tambe, A.~M. Rush, G.-Y. Wei,
  and D.~Brooks, ``Masr: A modular accelerator for sparse rnns,'' in \emph{2019
  28th International Conference on Parallel Architectures and Compilation
  Techniques (PACT)}.\hskip 1em plus 0.5em minus 0.4em\relax IEEE, 2019, pp.
  1--14.

\bibitem{parashar2017scnn}
A.~Parashar, M.~Rhu, A.~Mukkara, A.~Puglielli, R.~Venkatesan, B.~Khailany,
  J.~Emer, S.~W. Keckler, and W.~J. Dally, ``Scnn: An accelerator for
  compressed-sparse convolutional neural networks,'' \emph{ACM SIGARCH Computer
  Architecture News}, vol.~45, no.~2, pp. 27--40, 2017.

\bibitem{han2017ese}
S.~Han, J.~Kang, H.~Mao, Y.~Hu, X.~Li, Y.~Li, D.~Xie, H.~Luo, S.~Yao, Y.~Wang
  \emph{et~al.}, ``Ese: Efficient speech recognition engine with sparse lstm on
  fpga,'' in \emph{Proceedings of the 2017 ACM/SIGDA International Symposium on
  Field-Programmable Gate Arrays}, 2017, pp. 75--84.

\bibitem{dong2017more}
X.~Dong, J.~Huang, Y.~Yang, and S.~Yan, ``More is less: A more complicated
  network with less inference complexity,'' in \emph{Proceedings of the IEEE
  Conference on Computer Vision and Pattern Recognition}, 2017, pp. 5840--5848.

\bibitem{kim2018mosaic}
C.~Kim, D.~Shin, B.~Kim, and J.~Park, ``Mosaic-cnn: A combined two-step zero
  prediction approach to trade off accuracy and computation energy in
  convolutional neural networks,'' \emph{IEEE Journal on Emerging and Selected
  Topics in Circuits and Systems}, vol.~8, no.~4, pp. 770--781, 2018.

\bibitem{figurnov2015perforatedcnns}
M.~Figurnov, A.~Ibraimova, D.~Vetrov, and P.~Kohli, ``Perforatedcnns:
  Acceleration through elimination of redundant convolutions,'' \emph{arXiv
  preprint arXiv:1504.08362}, 2015.

\bibitem{mahmoud2018diffy}
M.~Mahmoud, K.~Siu, and A.~Moshovos, ``Diffy: A d{\'e}j{\`a} vu-free
  differential deep neural network accelerator,'' in \emph{2018 51st Annual
  IEEE/ACM International Symposium on Microarchitecture (MICRO)}.\hskip 1em
  plus 0.5em minus 0.4em\relax IEEE, 2018, pp. 134--147.

\bibitem{kligvasser2018xunit}
I.~Kligvasser, T.~R. Shaham, and T.~Michaeli, ``xunit: Learning a spatial
  activation function for efficient image restoration,'' in \emph{Proceedings
  of the IEEE Conference on Computer Vision and Pattern Recognition}, 2018, pp.
  2433--2442.

\end{thebibliography}

\end{document}